\documentclass[notitlepage,superscriptaddress,longbibliography,pra,twocolumn]{revtex4-1}
\usepackage{graphicx,amsfonts,amsmath,amssymb,mathrsfs,multirow,tikz,color}
\usepackage[mathcal]{euscript}
\usepackage[caption=false]{subfig}
\usepackage[colorlinks=true,citecolor=blue,urlcolor=purple]{hyperref}

\newcommand{\ket}[1]{\left| #1 \right\rangle}
\newcommand{\bra}[1]{\left\langle #1 \right|}
\newcommand{\ketbra}[2]{| #1 \rangle \langle #2 |}
\newcommand{\dprod}[2]{\left\langle #1, #2\right\rangle}
\newcommand{\abs}[1]{\left| #1\right|}
\newcommand{\mean}[1]{\langle #1 \rangle}

\newcommand{\CC}{\mathbb{C}}
\renewcommand{\d}{\mathrm{d}}
\newcommand{\me}{\mathrm{e}}
\newcommand{\FF}{\mathbb{F}}

\renewcommand{\i}{\mathrm{i}}

\newcommand{\SL}{\mathrm{SL}}

\newcommand{\I}{\openone}

\newcommand{\RR}{\mathbb{R}}
\newcommand{\X}{\mathcal{X}}

\newcommand{\ZZ}{\mathbb{Z}}

\renewcommand{\u}{\pmb{u}}
\renewcommand{\v}{\pmb{v}}

\renewcommand{\ge}{\geqslant}
\renewcommand{\geq}{\geqslant}
\renewcommand{\le}{\leqslant}
\renewcommand{\leq}{\leqslant}

\newcommand{\tr}{\operatorname{tr}}
\newcommand{\Tr}{\operatorname{Tr}}
\newcommand{\U}{\operatorname{U}}

\begin{document}

\title{Symmetries between measurements in quantum mechanics}

\author{H.~Chau Nguyen}
\email{chau.nguyen@uni-siegen.de}
\affiliation{Naturwissenschaftlich-Technische Fakult\"at, Universit\"at Siegen, Walter-Flex-Stra{\ss}e 3, 57068 Siegen, Germany}
\author{S\'ebastien Designolle}
\email{sebastien.designolle@unige.ch}
\affiliation{Group of Applied Physics, University of Geneva, 1211 Geneva, Switzerland}
\author{Mohamed Barakat}
\email{mohamed.barakat@uni-siegen.de}
\affiliation{Naturwissenschaftlich-Technische Fakult\"at, Universit\"at Siegen, Walter-Flex-Stra{\ss}e 3, 57068 Siegen, Germany}
\author{Otfried G\"uhne}
\email{otfried.guehne@uni-siegen.de}
\affiliation{Naturwissenschaftlich-Technische Fakult\"at, Universit\"at Siegen, Walter-Flex-Stra{\ss}e 3, 57068 Siegen, Germany}

\date{\today}

\begin{abstract}
  Symmetries are a key concept to connect mathematical elegance with physical insight.
  We consider measurement assemblages in quantum mechanics and show how their symmetry can be described by means of the so-called discrete bundles.
  It turns out that many measurement assemblages used in quantum information theory as well as for studying the foundations of quantum mechanics are entirely determined by symmetry; moreover, starting from a certain symmetry group, novel types of measurement sets can be constructed.
  The insight gained from symmetry allows us to easily determine whether the measurements in the set are incompatible under noisy conditions, i.e., whether they can be regarded as genuinely distinct ones.
  In addition, symmetry enables us to identify finite sets of measurements having a high sensitivity to reveal the quantumness of distributed quantum states.
\end{abstract}

\maketitle

\textit{Introduction.---}
Physics in all areas is alluded by symmetry.
Symmetry is at the heart of the understanding of crystals, lies at the foundation of general relativity, and sets the basis for modern quantum field theory.
In fact, Feynman considers symmetry as the main characteristics of the laws of physics~\cite{Feynman1965a}.

In quantum mechanics, measurements play a crucial role as they are the intermediate layer to transfer information from the `hidden' quantum mechanical world to the classical one.
Actually, one often works with several such measurements at the same time: state tomography~\cite{Nielsen2010a}, uncertainty relations~\cite{CBTW17}, quantum random access codes~\cite{ALMO08}, nonlocality~\cite{BCP+13}, quantum steering~\cite{Uola2019a}, or contextuality~\cite{Spe05, budroni2020} do all involve \emph{measurement~assemblages} with two or more measurements.
Consequently, understanding the relations among several measurements is crucial in quantum mechanics.

\begin{figure}[t]
  \subfloat[\label{sfig:1a}]{\includegraphics[width=0.18\textwidth]{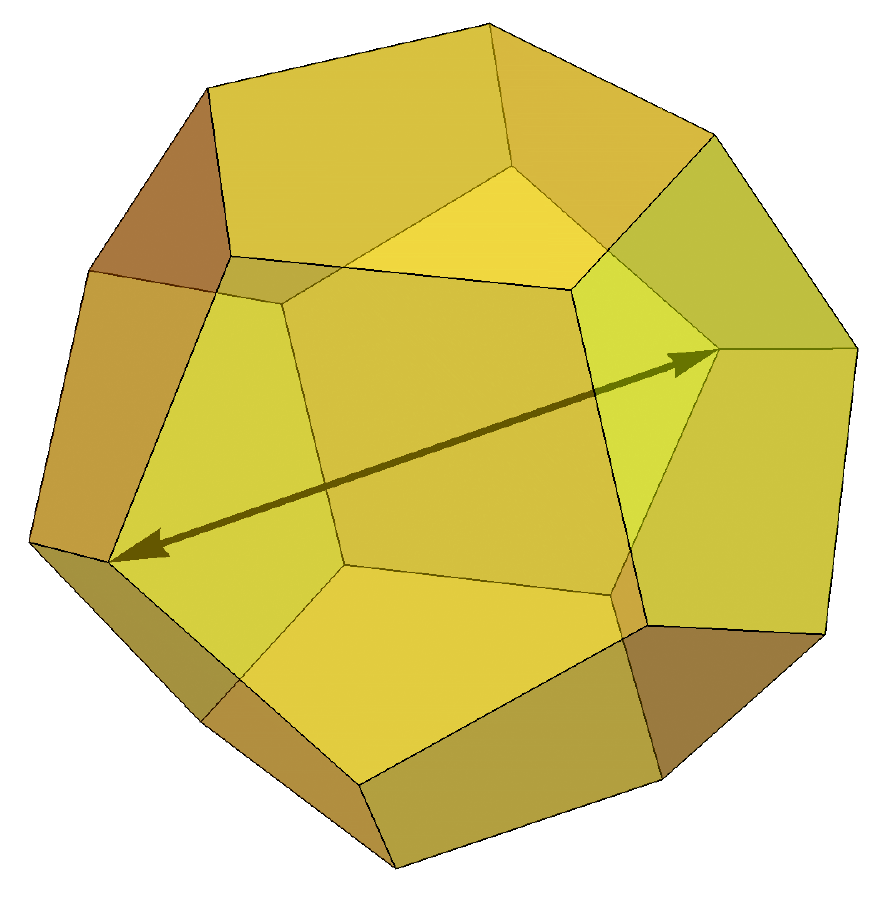}}
  \hspace{1cm}
  \subfloat[\label{sfig:1b}]{\includegraphics[width=0.20\textwidth]{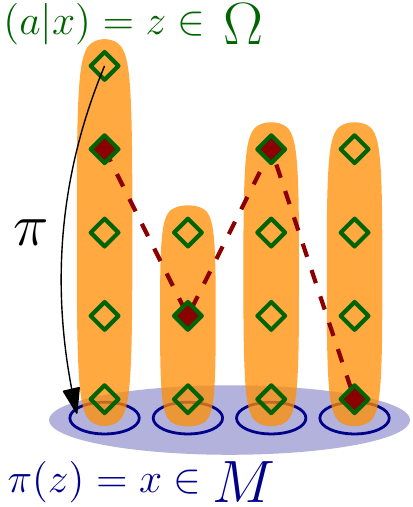}}
  \caption{
    (a)~The set of measurements on a qubit defined by the directions of the vertices of a regular dodecahedron.
    The two opposite arrows illustrate one of the ten projective measurements.
    (b)~Geometrical illustration of the bundle of measurement outcomes for $\abs{M}=4$ measurements (bottom ellipse).
    The diamonds denote the measurement outcomes, in total, $\abs{\Omega} = 16$, grouped vertically into fibres which have $5$, $3$, $4$, and $4$ outcomes (from left to right) corresponding to the $4$ measurements.
    The filled diamonds (one per measurement, connected by a dashed line) illustrate a section of the bundle.
  }
  \label{fig:1}
\end{figure}

To give a concrete example, for the experimental demonstration of the hierarchy of quantum correlations, measurements on a qubit along ten different directions have been used~\cite{SJWP10}, which form a dodecahedron (see Fig.~\ref{sfig:1a}).
Another instance is the standard construction of a complete set of mutually unbiased bases (MUBs)~\footnote{We use the term MUBs in a restrictive sense.
Specifically, MUBs in this paper always refer to the full set of $d+1$ mutually unbiased bases in dimension $d$ constructed by a specific standard procedure; see Ref.~\cite{DEBZ10} and Appendix~\ref{app:mub}.}.
Complete sets of MUBs play an important role in quantum information processing tasks such as quantum state tomography~\cite{Iva81,WF89}, quantum error correction~\cite{CRSS97}, entropic uncertainty relations~\cite{GM88}, or quantum key distribution~\cite{CBKG02}.

When talking about measurements, we are not working with physical entities at a single time, such as atoms in a crystal, the spacetime, or a quantum field, but rather with physical realisations of different measurements that cannot be carried out simultaneously.
Still, one can intuitively expect that the symmetry between the different measurements plays some important role.
Curiously, while the use of symmetry in the foundation of quantum mechanics and in quantum information theory was considered in several situations, see Refs.~\cite{Vollbrecht2001a,EMV04,RG06,SS16,Kiukas2017a,TRR19,Haapasalo2019a} to mention a few, a framework to describe the symmetry of a measurement assemblage is so far not available.

In this paper, we combine mathematical methods from group theory and the concept of discrete bundles with the physical description of measurements in quantum mechanics.
This results in a general approach to characterise the symmetry of a measurement assemblage.
So far, the concept of vector bundles has been widely used in physics, as they play a fundamental role in general relativity, gauge field theory and topological quantum matter~\cite{Baez1994a,Bernvevig2013a,Sharpe1997a}.
While discrete bundles may look a bit unfamiliar at first sight, they are conceptually simpler.

Using our methods we then identify a class of highly symmetric measurement assemblages for which not only the symmetry is specified by the measurements, but dually the measurements are determined by their symmetry.
The platonic assemblages (such as the above-mentioned dodecahedron) and MUBs are examples of such highly symmetric structures.
Conversely, starting from a symmetry group, we show that one can construct novel measurement assemblages that are highly symmetric.
Not only limited to qubits like the platonic assemblages, and more flexible than MUBs, these measurement assemblages can potentially find important applications.
As an illustration, the `quantumness' of these assemblages as characterised by their so-called incompatibility can be directly derived from their symmetry.
In addition, the obtained results allow us to demonstrate that some of the newly constructed assemblages, despite being finite, are more efficient in extracting quantum correlations than the whole infinite set of dichotomic measurements.
The existence of such a set has been pointed out in Ref.~\cite{Nguyen2019b}, but a concrete construction was not possible.

\textit{The bundle of measurement outcomes.---}
Consider a set of measurements labelled by $x \in M$ (see Fig.~\ref{sfig:1b}).
The outcomes of the measurement $x$ are conventionally denoted by $(a|x)$, which keeps track of both the outcome and the measurement it belongs to.
It is, however, convenient to separate the information by first introducing the (disjoint) union of all sets of outcomes $z=(a|x)$, denoted by $\Omega$.
The set $\Omega$ does not fully describe the outcomes of the measurements, since it lacks the information of which measurement the outcome belongs to.
Thus we introduce a map $\pi: \Omega \to M$ that projects the outcomes $z=(a|x)$ onto the corresponding measurement, that is, $\pi (z) = x$.
The triplet $(\Omega,\pi,M)$ is called a bundle.
Note that $\pi^{-1} (x)$, called the fibre over $x$, is precisely the set of the outcomes of the measurement $x$.
For concreteness, we assume that both $M$ and $\Omega$ are finite and the bundle is therefore discrete.

In quantum mechanics, to describe a measurement assemblage on a system of dimension $d$, one associates a so-called effect to every outcome $z$ in the bundle $\Omega$.
This is a positive semidefinite operator $A_z$, fulfiling the normalisation
\begin{equation}
  \sum_{z \in \pi^{-1} (x)} A_z = \I \mbox{ for all $x \in M$}.
  \label{eq:normalisation}
\end{equation}
Note that $\{A_z:z \in \pi^{-1} (x)\}$ is the set of effects of the measurement $x$ in the familiar terminology~\cite{Heinosaari2011a}.
This normalisation simply ensures that the probabilities for the outcomes of the measurement sum up to one.

\textit{Symmetry of measurement assemblages.---}
A possible symmetry of the assemblage $A$ may be described by a symmetry group $G$ together with a unitary representation $U:G \to \mathrm{U}(d)$ in the following way: the group $G$ permutes the outcomes $\Omega$ in a way that is compatible with the assignment of the outcomes to the measurements $M$,
\begin{equation}
  g [\pi (z)] = \pi [g(z)] \mbox{ for all $g \in G$ and $z\in\Omega$}.
  \label{eq:covariance}
\end{equation}
Moreover, measurement effects of different outcomes that are related by a symmetry element $g$ are also related by the corresponding unitary operator $U_g$,
\begin{equation}
  A_z = U_g A_{g^{-1}(z)} U^{-1}_g \mbox{ for all $g \in G$ and $z\in\Omega$},
  \label{eq:symmetry-assemblage}
\end{equation}
which may also be written as $A=g(A)$ for all $g\in G$.

\textit{The dodecahedron as an example.---}
The dodecahedron assemblage consists of $\abs{M}=10$ measurements on a qubit corresponding to ten lines connecting antipodal vertices of a regular dodecahedron (see Fig.~\ref{sfig:1a}).
Each fibre consists of two outcomes corresponding to two vertices lying on the same line (spin up or down).
The bundle of outcomes then contains $\abs{\Omega}=20$ points and the symmetry group consists of $60$ rotations~\footnote{For the symmetry group of the dodecahedron, see, e.g., Ref.~\cite{Sternberg1994a}.
Here we consider only rotations, but reflections can also be taken into account when one also allows for antiunitary representations.}.
Under these transformations, the different vertices are transformed into each other (action on the outcomes $\Omega$), and the different lines are also transformed into each other (action on the measurements $M$).
Crucially, the transformations respect the bundle structure by satisfying Eq.~\eqref{eq:covariance}, that is, the line connecting two rotated antipodal vertices is the same as the rotated image of the line connecting the two original antipodal vertices.
The assemblage then associates each vertex with a projection of the qubit onto that direction.
It is well-known that any rotation can be associated to a unitary transformation acting on the qubit~\cite{Sternberg1994a}.
Importantly, if vertices are transformed into each other, then the corresponding operators are also transformed into each other by the unitary operators according to Eq.~\eqref{eq:symmetry-assemblage}.

\textit{Uniform and rigidly symmetric assemblages.---}
There are two properties of the dodecahedron assemblage that
are worth to point out.
Firstly, for this assemblage, any outcome can be related to any other by a symmetry transformation.
In this case, all outcomes are in fact equivalent; we say the assemblage is \emph{uniform}.

Secondly, let us pick a vertex $z$ and consider all rotational symmetries of the dodecahedron that leave this point invariant.
This is known as the stabiliser (sub)group of that vertex, denoted by $G_z$.
With Eq.~\eqref{eq:symmetry-assemblage} it is then clear that the corresponding effect $A_z$ commutes with all the unitary operators of the stabiliser group $U(G_z)=\{U_g:g\in G_z\}$.
For the dodecahedron, the only projections commuting with all of the unitary operators from the stabiliser group at a vertex are in fact (i)~the spin projection in the direction of the vertex and (ii)~its complement.

In general, if for all outcomes $z$ the set of all operators that commute with the stabiliser $U(G_z)$ is spanned by (i)~a single projection $\Pi_z$ and (ii)~its complement $\I-\Pi_z$, we say that the symmetry is \emph{rigid}.
The only two ways for a rigidly symmetric assemblage $A$ to be projective are either $A_z=\Pi_z$ or $A_z=\I-\Pi_z$.
In this sense, we say that the assemblage is determined by its symmetry.
By representation theory of groups, this is equivalent to saying that the representation $U$ restricted to $G_z$ contains exactly \emph{two} irreducible subrepresentations, which can be easily verified by character theory~\cite{Serre1977a}.

All platonic assemblages for qubits are easily seen to be uniform and rigidly symmetric (see also below).
Later we will also demonstrate that MUBs arise from uniform and rigid symmetries.
We further show that such uniform and rigidly symmetric assemblages can be systematically constructed from chosen symmetry groups and their representations; see Appendix~\ref{app:construction}.
There we illustrate this procedure with the so-called finite complex reflection groups~\cite{ST54}, which are already used in the context of complex projective designs~\cite{BW13,HW18}.
Here we show that they also allow for the construction of various uniform and rigidly symmetric measurement assemblages enumerated in Table~\ref{tab:marvel}.

\begin{table*}[ht!]
  \centering
  \begin{tabular}{|c|c|c|c|c|c|c|c|}
    \hline
    $~d~$                  & ~Group~                & $\abs{M}$      & Comments            & $\quad\alpha^\ast=\max\{\eta:A^\eta\ \mathrm{compatible}\}\quad$ & $\quad\beta^\ast=\max\{\eta:\bar{A}^\eta\ \mathrm{compatible}\}\quad$ \\ \hline
    \multirow{6}{*}{2}     & \multirow{3}{*}{ST 8}  & 3              & Octahedron --- MUBs & \multicolumn{2}{c|}{$\frac{1}{\sqrt{3}}\approx0.5774$ \cite{Bus86,SJWP10,ULMH16}}                                                        \\ \cline{3-6}
                           &                        & 4              & Cube                & \multicolumn{2}{c|}{$\frac{1}{\sqrt{3}}\approx0.5774$ \cite{SJWP10}}                                                                     \\ \cline{3-6}
                           &                        & 6              & Cuboctahedron       & \multicolumn{2}{c|}{$\frac13\sqrt{\frac52}\approx0.5270$}                                                                                \\ \cline{2-6}
                           & \multirow{3}{*}{ST 16} & 6              & Icosahedron         & \multicolumn{2}{c|}{$\frac{1+\sqrt{5}}{6}\approx0.5393$ \cite{SJWP10,ULMH16}}                                                            \\ \cline{3-6}
                           &                        & 10             & Dodecahedron        & \multicolumn{2}{c|}{$\frac{3+\sqrt{5}}{10}\approx0.5236$ \cite{SJWP10,ULMH16}}                                                           \\ \cline{3-6}
                           &                        & 15             & Icosidodecahedron   & \multicolumn{2}{c|}{$\frac{\sqrt{31+12\sqrt{5}}}{15}\approx0.5070$}                                                                      \\ \hline
    \multirow{4}{*}{3}     & ST 24                  & 7              &                     & $\approx0.4960$                                                  & $\approx0.7556$                                                       \\ \cline{2-6}
                           & ST 25                  & 4              & MUBs                & $\frac{1+3\sqrt{5}}{16}\approx0.4818$ \cite{DSFB19}              & 1 \cite{Skrzypczyk2014a}                                              \\ \cline{2-6}
                           & \multirow{2}{*}{ST 27} & 15             &                     & $\frac{3+\sqrt{5}+\sqrt{94+30\sqrt{5}}}{40}\approx0.4482$        & $\frac{\sqrt{5}+\sqrt{75+30\sqrt{5}}}{20}\approx0.7078^\ddag$         \\ \cline{3-6}
                           &                        & 20             &                     & $\approx0.4443$                                                  & $\frac{5+3\sqrt{5}+\sqrt{6(189+65\sqrt{5})}}{80}\approx0.7062^\ddag$  \\ \hline
    \multirow{7}{*}{4}     & ST 28                  & 3              & Real MUBs           & $\frac59\approx0.5556$                                           & 1                                                                     \\ \cline{2-6}
                           & \multirow{3}{*}{ST 29} & 5              & MUBs                & $\frac{3+2\sqrt{3}}{15}\approx0.4309$ \cite{DSFB19}              & $\frac{\sqrt{5}+\sqrt{10-2\sqrt{5}}}{5}\approx0.9174$                 \\ \cline{3-6}
                           &                        & 10             &                     & $\approx0.4167$                                                  & $\approx0.8857$                                                       \\ \cline{3-6}
                           &                        & 20             &                     & $\gtrapprox0.4107$                                               & $\gtrapprox0.8143$                                                    \\ \cline{2-6}
                           & ST 30                  & 75             &                     & $\gtrapprox0.4947$                                               & $\gtrapprox0.8874$                                                    \\ \cline{2-6}
                           & \multirow{2}{*}{ST 31} & 15             &                     & $\frac{7+2\sqrt{31}}{45}\approx0.4030$                           & $\frac{\sqrt{5}+\sqrt{50+22\sqrt{5}}}{15}\approx0.8130^\ddag$         \\ \cline{3-6}
                           &                        & 120            &                     & $\gtrapprox0.3553$                                               & $\gtrapprox0.7672$                                                    \\ \hline
  \end{tabular}
  \caption{
    Projective measurement assemblages constructed from the finite complex reflection groups and their incompatibility (see Appendix~\ref{app:construction} for the details of the construction).
    The number $d$ is the dimension, the groups are given by their Shephard--Todd (ST) number~\cite{ST54}, and $\abs{M}$ is the number of measurements.
    The last two columns illustrate the insight {gained} from symmetry by giving analytically two interesting incompatibility properties of the assemblages: $\alpha^\ast$ and $\beta^\ast$ as defined in the text.
    Their values can all be exactly represented (with radicals), but large representations are converted into numeric values.
    Equivalently, the quantities $\alpha^\ast$ and $\beta^\ast$ correspond to the noise threshold for steering of the isotropic (left) and the Werner (right) states (see Appendix~\ref{app:steering}).
    The symbol $\ddag$ indicates the (finite) projective measurement assemblages performing better than the infinite set of all two-outcome measurements~\cite{Nguyen2019b}.
    For too large $\abs{M}$, only bounds on $\alpha^\ast$ and $\beta^\ast$ can be obtained, thanks to a heuristic method inspired by statistical mechanics (see Appendix~\ref{app:rigid-symmetry}).
    Note that many groups are not represented as they give equivalent measurement assemblages.
  }
  \label{tab:marvel}
\end{table*}

\textit{Symmetry and measurement incompatibility.---}
Before going more into the detailed analysis of the symmetry of the assemblages, let us illustrate how we can use the symmetry to easily analyse, for instance, the incompatibility of measurement assemblages~\cite{Heinosaari2011a}.
Determining the incompatibility of an assemblage is fundamental in quantum mechanics and in many quantum information applications because measurements in a compatible assemblage, despite appearing as distinct, can in fact be derived from a single \emph{parent measurement}.
As such, they cannot actually provide advantage in various quantum phenomena such as uncertainty relations~\cite{CBTW17}, random access codes~\cite{CHT20}, or Bell inequalities~\cite{WPF09}.

Let us introduce one more necessary mathematical concept to deal with the concept of incompatibility: the sections of the bundle.
A section $s$ of the bundle $\Omega$ is a map $s:M\to\Omega$ such that $\pi[s(x)]=x\in M$.
Intuitively, it is a choice of one outcome from each measurement (see Fig.~\ref{sfig:1b}).
The set of all sections of $\Omega$ is denoted by $\Gamma(\Omega)$.
The measurement assemblage $A$ is said to be compatible if there is a parent measurement with output in $\Gamma(\Omega)$ such that
\begin{equation}
  A_z = \!\!\!\sum_{s\in\Gamma(\Omega)} \delta_{s[\pi(z)],z} F_s.
  \label{eq:compatibility-def}
\end{equation}
One can easily verify that this reduces to the usual definition of incompatibility of a finite measurement assemblage such as in~Ref.~\cite{Heinosaari2011a}.

In reality, it is necessary to consider the imperfections of the measurements due to noise.
As a simple model of the noise, one can consider the white noise acting on the assemblage, leading to a noisy one $A^\eta_z=\eta A_z + {(1-\eta) \Tr(A_z) \I/d}$ with $0 \le \eta \le 1$.
One can ask up to which level of noise the assemblage remains incompatible,
\begin{equation}
  \alpha^\ast = \max\{\eta\leq1:\mbox{$A^\eta$ is compatible}\}.
  \label{eq:primal}
\end{equation}
For specificity, we focus the discussion on this white noise and present results also for another type of noise, $\beta^\ast = \max\{\eta\leq1:\mbox{$\bar{A}^\eta$ is compatible}\}$, with $\bar{A}^\eta_z = {\eta(\Tr(A_z)\I-A_z)/(d-1)} + {(1-\eta) \Tr(A_z)\I/d}$.
The reason for our choice is motivated by an application of measurement incompatibility in quantum steering~\cite{Quintino2014a,UBGP15}.
More precisely, the quantity $\alpha^\ast$ (resp.~$\beta^\ast$) corresponds to the visibility from which steering can be demonstrated with the isotropic (resp.~Werner) state (see Appendix~\ref{app:steering} for details).
The reader should note, however, that all of our discussion can be adapted to consider other types of noise such as those considered in Ref.~\cite{DFK19}.

Computing the noise thresholds $\alpha^\ast$ and $\beta^\ast$ can be done via semidefinite programming (SDP)~\cite{BV04}.
However, the number of variables in the problem grows as $\abs{\Gamma(\Omega)}$, which is exponential in the number $\abs{M}$ of measurements and thus makes it quickly intractable.
Here we illustrate that for a uniform and rigidly symmetric assemblage the insight from symmetry allows one to derive rather explicit formulae for $\alpha^\ast$ and $\beta^\ast$, even when the original SDPs are intractable.

Although the analysis of the symmetry of the SDP~\eqref{eq:primal} can be carried out (see Appendix~\ref{app:simplification}), deeper insight can be gained when approaching the problem from the dual perspective~\cite{BV04}.
In this case, duality theory implies that $\alpha^\ast$ can be computed by an equivalent dual problem,
\begin{align}
  \label{eq:dual}
  \alpha^\ast =\min_X \quad & 1 + \sum_{z \in \Omega} \Tr(X_z A_z) \\
  \mbox{s.t.} \quad & 1 + \sum_{z \in \Omega} \Tr(X_z A_z)
  \ge \frac{1}{d} \sum_{z \in \Omega} \Tr(A_z) \Tr(X_z) \nonumber \\
  & \sum_{z \in \Omega} \delta_{s[\pi(z)],z} X_z \ge 0
  \qquad \forall s \in \Gamma(\Omega).
  \nonumber
\end{align}
Note that the dual  variable $X$ is associated to every outcome $z$ exactly like $A$.
Now, when $A$ is symmetric under $G$, a standard argument from group theory allows us to impose that $X$ is also symmetric under $G$ in the same way as Eq.~\eqref{eq:symmetry-assemblage}, namely, $X_z= U_g X_{g^{-1}(z)} U_{g}^{-1}$ (see Appendix~\ref{app:simplification} for details).
This implies that $X_z$, like $A_z$, commutes with all of the stabiliser $U(G_z)$.
In particular, if the assemblage $A$ is uniform and rigidly symmetric, then the only possibility is $X_z=a\I+b A_z$.
Note that $a$ and $b$ also do not depend on the particular outcome $z$ because all outcomes are equivalent for uniform assemblages.
Interestingly, such a form of the solution has been used as an ad hoc ansatz in Ref.~\cite{DSFB19}.
While case by case inspections could sometimes demonstrate its optimality~\cite{DSFB19,SJWP10,ULMH16,DFK19}, here we see that this ansatz as well as its optimality are in fact simple consequences of the symmetry of the assemblage.
This allows us to systematically reorganise known results that were scattered in the literature and to easily derive the quantities $\alpha^\ast$ and $\beta^\ast$ for many other symmetric assemblages; see Table~\ref{tab:marvel}.
The procedure for fixing the parameters $a$ and $b$ together with the explicit formulae for $\alpha^\ast$ and $\beta^\ast$ are given in Appendix~\ref{app:rigid-symmetry}.

We would like to emphasise two interesting consequences of our results on incompatibility.
First, there are some newly constructed measurement assemblages
(indicated by $\ddag$ in Table~\ref{tab:marvel}) having three or four outcomes, which are more incompatible than the set of \emph{all} measurements with two outcomes (dichotomic measurements) in the same dimension.
This means that they can reveal quantum steering in a situation where all dichotomic measurements cannot~\cite{Nguyen2019b}.
Secondly, MUBs in odd prime power dimensions cannot be used to steer the Werner state (see Appendix~\ref{app:steering} for details), which generalises the numerical result obtained in dimension three in Ref.~\cite{Skrzypczyk2014a}.

\textit{Determination of uniformity and rigidity.---}
The problem of determining and investigating the symmetry of a measurement assemblage is interesting in its own right.
The symmetry groups of the assemblages defined by the platonic solids are in fact special cases of the complex reflection groups as visible in Table~\ref{tab:marvel}.
Their uniformity and rigidity follow then directly from the construction.
Let us show that MUBs are also uniform and rigidly symmetric.
For a quantum system of prime power dimension~$d$, there is a standard construction of $d+1$ rank-one projective measurements where the effects from different measurements have exactly the same overlap of $1/\sqrt{d}$~\cite{DEBZ10}, which have been referred to as MUBs throughout this paper.
For concreteness, we sketch the argument below only for odd prime dimensions; a general proof valid for any prime power dimensions is given in Appendix~\ref{app:mub}.

\begin{figure}[!t]
  \subfloat[\label{sfig:2a}]{\includegraphics[height=3cm]{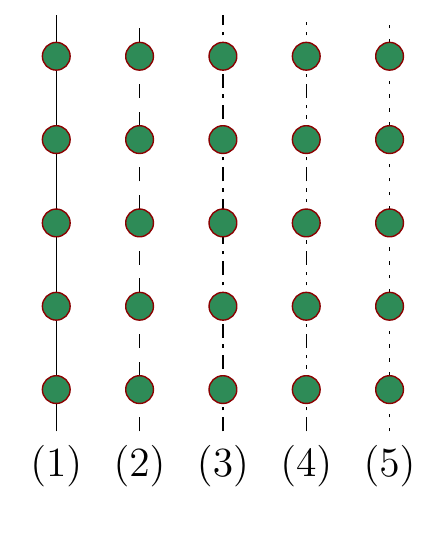}}
  \subfloat[\label{sfig:2b}]{\includegraphics[height=3cm]{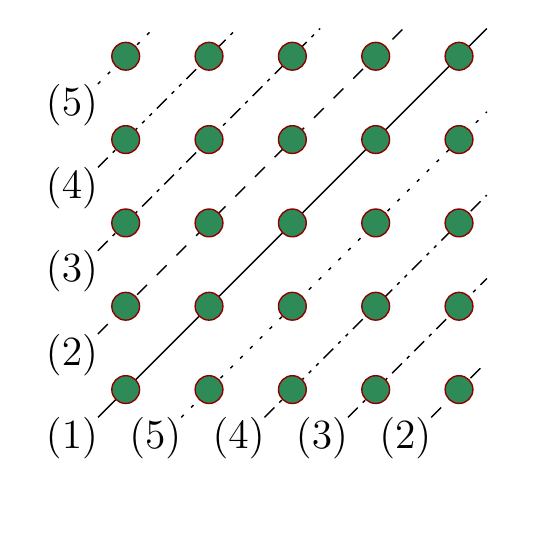}}
  \subfloat[\label{sfig:2c}]{\includegraphics[height=3cm]{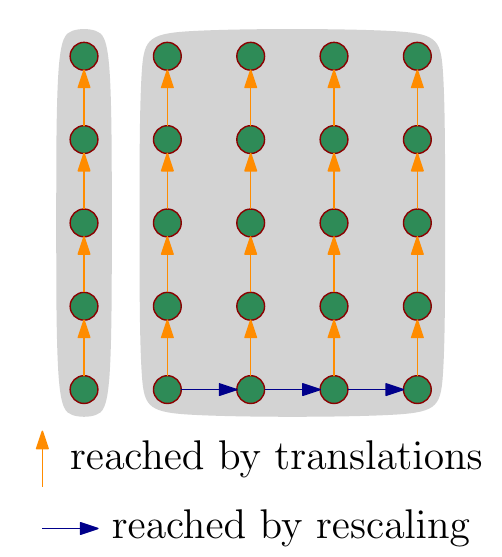}}
  \caption{
    The phase space for a quantum system of dimension $d$ (here $d=5$) is the plane $(\ZZ_d)^2$.
    (a) The striation of the phase space into five vertical lines corresponding to the measurement in the computational basis.
    (b) Another striation of the phase space $(\ZZ_d)^2$ corresponding to another measurement mutually unbiased to the previous one.
    Lines are numbered from $(1)$ to $(5)$.
    (c) Orbits of the stabiliser group of the vertical axis.
    Vertical translations imply that all points on vertical lines are in the same orbit (vertical arrows).
    Rescaling the two axes by two opposite scaling factors implies that all points on the horizontal axis, except for the origin, are in the same orbit (horizontal arrows).
  }
  \label{fig:2}
\end{figure}

The symmetry of MUBs and their rigidity can be elegantly seen in the discrete phase space representation~\cite{Woo87}.
A quantum system can be represented by a two-dimensional discrete phase space $(\ZZ_d)^2$~\cite{Woo87}, where $\ZZ_d$ denotes the field of integer residual classes of the prime divisor $d$.
A measurement in one of the MUBs corresponds to a striation of the plane, that is, a partition of the plane into parallel lines.
For example, vertical lines correspond to projections onto the computational basis, see Fig.~\ref{sfig:2a}.
Similarly Fig.~\ref{sfig:2b} illustrates another measurement in one of the MUBs corresponding to another striation.
There are exactly $d+1$ such striations forming $d+1$ measurements in the MUBs.

The symmetry of MUBs can be described by linear translations and linear transformations with unit determinant over the phase space $(\ZZ_d)^2$~\cite{App05,App09}.
Clearly these transformations allow one to transform any line into any other, thus establishing the uniformity of MUBs.
Moreover, the rigidity condition amounts to the stabiliser group of a line having exactly two orbits, one of which being the line itself, and the other its complement (see Appendix~\ref{app:mub} for the details).
Since all lines are equivalent (uniformity), we can consider the vertical axis for specificity.
All linear translations parallel to the axis clearly leave it invariant, thus are in the stabiliser group of the axis.
Moreover, rescaling the two axes with opposite scaling factors also leave the axis invariant.
It is then straightforward to see that the stabiliser group has indeed exactly two orbits, the axis itself and its complement as illustrated in Fig.~\ref{sfig:2c}.

\textit{Conclusion.---}
We have demonstrated how the symmetry of a set of several measurements can be formalised by means of discrete bundles.
Determining the symmetry groups for various assemblages, we have also shown how insightful conclusions can be drawn from their symmetry.
Further study of the symmetry of other measurement assemblages such as MUBs with non standard construction, or incomplete sets of MUBs, could shed light on their nature.
Starting from suitable symmetry groups, we have constructed new measurement assemblages with novel properties and analysed some of these properties.
More detailed analysis and further applications of these measurements in quantum information processing are to be expected in the future; for this purpose, we make them available online~(see Appendix~\ref{app:sqma}).
More broadly, in addition to works in different contexts~\cite{TG20,TRR19,SS16}, we believe that further analysis of symmetry of different protocols will significantly deepen our understanding of other topics of the foundations of quantum mechanics and quantum information theory.

\begin{acknowledgements}
  We thank Marcus Appleby, Johannes Berg, Nicolas Brunner, Jonathan Steinberg, Roope Uola, and Shayne Waldron for fruitful discussions.
  This work was supported by the DFG and the ERC (Consolidator Grant 683107/TempoQ).
  Financial support by the Swiss National Science Foundation (Starting grant DIAQ, NCCR-QSIT) is acknowledged.
  HCN thanks the VNUHCM Center for Defense and Security Training for giving him a two-week accomodation.
\end{acknowledgements}

\bibliography{quantum-steering}

\begin{thebibliography}{10}

\bibitem{Feynman1965a}
R.~Feynman, {\em The character of physical laws}.
\newblock The MIT Press, 1965.

\bibitem{Nielsen2010a}
M.~A. Nielsen and I.~L. Chuang, {\em Quantum computation and quantum
  information}.
\newblock Cambridge University Press, 2010.

\bibitem{CBTW17}
P.~J. Coles, M.~Berta, M.~Tomamichel, and S.~Wehner, ``Entropic uncertainty
  relations and their applications,'' {\em Rev. Mod. Phys.}, vol.~89,
  p.~015002, 2017.

\bibitem{ALMO08}
A.~Ambainis, D.~Leung, L.~Mancinska, and M.~Ozols, ``Quantum random access
  codes with shared randomness,'' {\em arXiv:0810.2937}, 2008.

\bibitem{BCP+13}
N.~Brunner, D.~Cavalcanti, S.~Pironio, V.~Scarani, and S.~Wehner, ``Bell
  nonlocality,'' {\em Rev. Mod. Phys.}, vol.~86, pp.~419--478, 2014.

\bibitem{Uola2019a}
R.~Uola, A.~C.~S. Costa, H.~C. Nguyen, and O.~G\"uhne, ``Quantum steering,''
  {\em Rev. Mod. Phys.}, vol.~92, no.~015001, 2019.

\bibitem{Spe05}
R.~W. Spekkens, ``Contextuality for preparations, transformations, and unsharp
  measurements,'' {\em Phys. Rev. A}, vol.~71, p.~052108, 2005.

\bibitem{budroni2020}
C.~Budroni, A.~Cabello, O.~G\"uhne, M.~Kleinmann, and J.-A. Larsson, ``Quantum
  contextuality,'' 2020.
\newblock in preparation.

\bibitem{SJWP10}
D.~J. Saunders, S.~J. Jones, H.~M. Wiseman, and G.~J. Pryde, ``{Experimental
  EPR-steering using Bell-local states},'' {\em Nature Physics}, vol.~6,
  no.~11, pp.~845--849, 2010.

\bibitem{Note1}
We use the term MUBs in a restrictive sense. Specifically, MUBs in this paper
  always refer to the full set of $d+1$ mutually unbiased bases in dimension
  $d$ constructed by a specific standard procedure; see Ref.~\cite {DEBZ10} and
  Appendix~\ref {app:mub}.

\bibitem{Iva81}
I.~D. Ivanovic, ``Geometrical description of quantal state determination,''
  {\em J. Phys. A: Math. and Gen.}, vol.~14, no.~12, pp.~3241--3245, 1981.

\bibitem{WF89}
W.~K. Wootters and B.~D. Fields, ``Optimal state-determination by mutually
  unbiased measurements,'' {\em Ann. Phys.}, vol.~191, no.~2, pp.~363 -- 381,
  1989.

\bibitem{CRSS97}
A.~R. Calderbank, E.~M. Rains, P.~W. Shor, and N.~J.~A. Sloane, ``Quantum error
  correction and orthogonal geometry,'' {\em Phys. Rev. Lett.}, vol.~78,
  pp.~405--408, 1997.

\bibitem{GM88}
H.~Maassen and J.~B.~M. Uffink, ``Generalized entropic uncertainty relations,''
  {\em Phys. Rev. Lett.}, vol.~60, pp.~1103--1106, 1988.

\bibitem{CBKG02}
N.~J. Cerf, M.~Bourennane, A.~Karlsson, and N.~Gisin, ``Security of quantum key
  distribution using $\mathit{d}$-level systems,'' {\em Phys. Rev. Lett.},
  vol.~88, p.~127902, 2002.

\bibitem{Vollbrecht2001a}
K.~G.~H. Vollbrecht and R.~F. Werner, ``Entanglement measures under symmetry,''
  {\em Phys. Rev. A}, vol.~64, 2001.

\bibitem{EMV04}
Y.~C. {Eldar}, A.~{Megretski}, and G.~C. {Verghese}, ``Optimal detection of
  symmetric mixed quantum states,'' {\em IEEE Trans. Inf. Theory}, vol.~50,
  no.~6, pp.~1198--1207, 2004.

\bibitem{RG06}
J.~M. Renes and M.~Grassl, ``Generalized decoding, effective channels, and
  simplified security proofs in quantum key distribution,'' {\em Phys. Rev. A},
  vol.~74, p.~022317, Aug 2006.

\bibitem{SS16}
W.~Slomczynski and A.~Szymusiak, ``Highly symmetric {POVMs} and their
  informational power,'' {\em Quantum Information Processing}, vol.~15, no.~1,
  pp.~565--606, 2016.

\bibitem{Kiukas2017a}
J.~Kiukas, C.~Budroni, R.~Uola, and J.-P. Pellonp\"a\"a, ``Continuous-variable
  steering and incompatibility via state-channel duality,'' {\em Phys. Rev. A},
  vol.~96, p.~042331, 2017.

\bibitem{TRR19}
A.~Tavakoli, D.~Rosset, and M.-O. Renou, ``Enabling computation of correlation
  bounds for finite-dimensional quantum systems via symmetrization,'' {\em
  Phys. Rev. Lett.}, vol.~122, p.~070501, 2019.

\bibitem{Haapasalo2019a}
E.~Haapasalo, ``Compatibility of covariant quantum channels with emphasis on
  weyl symmetry,'' {\em Annales Henri Poincar{\'e}}, vol.~20, no.~9,
  pp.~3163--3195, 2019.

\bibitem{Baez1994a}
J.~Baez and J.~P. Muniain, {\em Gauge fields, knots and gravity}.
\newblock World Scientific, 1994.

\bibitem{Bernvevig2013a}
B.~A. Bernvevig and T.~L. Hughes, {\em Topological insulators and topological
  superconductors}.
\newblock Princeton University Press, 2013.

\bibitem{Sharpe1997a}
R.~W. Sharpe, {\em Differential geometry}.
\newblock Springer, 1997.

\bibitem{Nguyen2019b}
H.~C. Nguyen and O.~G\"uhne, ``Some quantum measurements with three outcomes
  can reveal nonclassicality where all two-outcome measurements fail,'' {\em
  arXiv:2001.03514}, 2020.

\bibitem{Heinosaari2011a}
T.~Heinosaari and M.~Ziman, {\em The mathematical language of quantum theory:
  from uncertainty to entanglement}.
\newblock Cambridge University Press, 2011.

\bibitem{Note2}
For the symmetry group of the dodecahedron, see, e.g., Ref.~\cite
  {Sternberg1994a}. Here we consider only rotations, but reflections can also
  be taken into account when one also allows for antiunitary representations.

\bibitem{Sternberg1994a}
S.~Sternberg, {\em Group theory and physics}.
\newblock Cambridge University Press, 1994.

\bibitem{Serre1977a}
J.-P. Serre, {\em Linear representation of finite groups}.
\newblock Springer, 1977.

\bibitem{ST54}
G.~C. Shephard and J.~A. Todd, ``Finite unitary reflection groups,'' {\em Can.
  J. Math.}, vol.~6, pp.~274--304, 1954.

\bibitem{BW13}
H.~Broome and S.~Waldron, ``On the construction of highly symmetric tight
  frames and complex polytopes,'' {\em Lin. Alg. App.}, vol.~439, no.~12,
  pp.~4135 -- 4151, 2013.

\bibitem{HW18}
D.~Hughes and S.~Waldron, ``Spherical (t,t)-designs with a small number of
  vectors,'' 2018.

\bibitem{Bus86}
P.~Busch, ``Unsharp reality and joint measurements for spin observables,'' {\em
  Phys. Rev. D}, vol.~33, pp.~2253--2261, 1986.

\bibitem{ULMH16}
R.~Uola, K.~Luoma, T.~Moroder, and T.~Heinosaari, ``Adaptive strategy for joint
  measurements,'' {\em Phys. Rev. A}, vol.~94, p.~022109, 2016.

\bibitem{DSFB19}
S.~Designolle, P.~Skrzypczyk, F.~Fr\"owis, and N.~Brunner, ``Quantifying
  measurement incompatibility of mutually unbiased bases,'' {\em Phys. Rev.
  Lett.}, vol.~122, p.~050402, 2019.

\bibitem{Skrzypczyk2014a}
P.~Skrzypczyk, M.~Navascu\'{e}s, and D.~Cavalcanti, ``Quantifying
  {Einstein-Podolsky-Rosen} steering,'' {\em Phys. Rev. Lett.}, vol.~112,
  p.~180404, 2014.

\bibitem{CHT20}
C.~Carmeli, T.~Heinosaari, and A.~Toigo, ``Quantum random access codes and
  incompatibility of measurements,'' {\em arXiv:1911.04360}, 2019.

\bibitem{WPF09}
M.~M. Wolf, D.~Perez-Garcia, and C.~Fernandez, ``Measurements incompatible in
  quantum theory cannot be measured jointly in any other no-signaling theory,''
  {\em Phys. Rev. Lett.}, vol.~103, p.~230402, 2009.

\bibitem{Quintino2014a}
M.~T. Quintino, T.~V\'{e}rtesi, and N.~Brunner, ``Joint measurability,
  {Einstein-Podolsky-Rosen} steering, and {Bell} nonlocality,'' {\em Phys. Rev.
  Lett.}, vol.~113, p.~160402, 2014.

\bibitem{UBGP15}
R.~Uola, C.~Budroni, O.~G\"uhne, and J.-P. Pellonp\"a\"a, ``One-to-one mapping
  between steering and joint measurability problems,'' {\em Phys. Rev. Lett.},
  vol.~115, p.~230402, 2015.

\bibitem{DFK19}
S.~Designolle, M.~Farkas, and J.~Kaniewski, ``Incompatibility robustness of
  quantum measurements: a unified framework,'' {\em New J. Phys.}, 2019.

\bibitem{BV04}
S.~Boyd and L.~Vandenberghe, {\em Convex Optimization}.
\newblock New York, NY, USA: Cambridge University Press, 2004.

\bibitem{DEBZ10}
T.~Durt, B.-G. Englert, I.~Bengtsson, and K.~\.Zyczkowski, ``On mutually
  unbiased bases,'' {\em Int. J. Quantum Inf.}, vol.~8, no.~4, pp.~535--640,
  2010.

\bibitem{Woo87}
W.~K. Wootters, ``A {W}igner-function formulation of finite-state quantum
  mechanics,'' {\em Ann. Phys.}, vol.~176, no.~1, pp.~1 -- 21, 1987.

\bibitem{App05}
D.~M. Appleby, ``{Symmetric informationally complete--positive operator valued
  measures and the extended Clifford group},'' {\em J. Math. Phys.}, vol.~46,
  no.~5, p.~052107, 2005.

\bibitem{App09}
D.~M. Appleby, ``{Properties of the extended Clifford group with applications
  to SIC-POVMs and MUBs},'' {\em arXiv:0909.5233}, 2009.

\bibitem{TG20}
A.~Tavakoli and N.~Gisin, ``The platonic solids and fundamental tests of
  quantum mechanics,'' {\em arXiv:2001.00188}, 2020.

\bibitem{Armstrong2010a}
M.~A. Armstrong, {\em Groups and symmetry}.
\newblock Springer, 2010.

\bibitem{Wal18}
S.~Waldron, {\em An introduction to finite tight frames}.
\newblock New York, NY: Birkh\"{a}user, 2018.

\bibitem{CGGZ19}
J.~Czartowski, D.~Goyeneche, M.~Grassl, and K.~\.Zyczkowski, ``Iso-entangled
  mutually unbiased bases, symmetric quantum measurements and mixed-state
  designs,'' {\em arXiv:1906.12291}, 2019.

\bibitem{Mezard2009a}
M.~M\'ezard and A.~Montanari, {\em Information, physics, and computation}.
\newblock Oxford University Press, 2009.

\bibitem{Press1992a}
W.~H. Press, S.~A. Teukolsky, W.~T. Vetterling, and B.~P. Flannery, {\em
  Numerical Recipes in C (2nd Ed.): The Art of Scientific Computing}.
\newblock Cambridge University Press, 1992.

\bibitem{Dotsenko1994a}
V.~Dotsenko, {\em An introduction to the theory of spin glasses and neural
  networks}.
\newblock World Scientific, 1994.

\bibitem{Goldenfeld1972a}
N.~Goldenfeld, {\em Lectures on phase transitions and the renormalization
  group}.
\newblock Addison-Wesley, 1972.

\bibitem{Wiseman2007a}
H.~M. Wiseman, S.~J. Jones, and A.~C. Doherty, ``Steering, entanglement,
  nonlocality, and the {Einstein-Podolsky-Rosen} paradox,'' {\em Phys. Rev.
  Lett.}, vol.~98, p.~140402, 2007.

\bibitem{CS16b}
D.~Cavalcanti and P.~Skrzypczyk, ``Quantum steering: a review with focus on
  semidefinite programming,'' {\em Rep. Prog. Phys.}, vol.~80, no.~2,
  p.~024001, 2016.

\bibitem{QVB14}
M.~T. Quintino, T.~V\'ertesi, and N.~Brunner, ``Joint measurability,
  {{Einstein--Podolsky--Rosen}} steering, and {{Bell}} nonlocality,'' {\em
  Phys. Rev. Lett.}, vol.~113, p.~160402, 2014.

\bibitem{Werner1989a}
R.~F. Werner, ``Quantum states with {Einstein-Podolsky-Rosen} correlations
  admitting a hidden-variable model,'' {\em Phys. Rev. A}, vol.~40, p.~4277,
  1989.

\bibitem{Durbin2009a}
J.~R. Durbin, {\em Mordern algebra: an introduction}.
\newblock John Wiley \& Son, 2009.

\bibitem{Gibbons2004a}
K.~S. Gibbons, M.~J. Hoffman, and W.~K. Wootters, ``Discrete phase space based
  on finite fields,'' {\em Phys. Rev. A}, vol.~70, p.~062101, 2004.

\bibitem{Bandyopadhyay2002a}
S.~Bandyopadhyay, P.~O. Boykin, V.~Roychowdhury, and F.~Vatan, ``A new proof
  for the existence of mutually unbiased bases,'' {\em Algorithmica}, vol.~34,
  p.~512, 2002.

\bibitem{App07}
D.~M. Appleby, ``Symmetric informationally complete measurements of arbitrary
  rank,'' {\em Opt. Spectrosc.}, vol.~103, no.~3, pp.~416--428, 2007.

\bibitem{Gottesman1998a}
D.~Gottesman, ``Theory of fault-tolerant quantum computation,'' {\em Phys. Rev.
  A}, vol.~57, pp.~127--137, 1998.

\bibitem{Wootters2007a}
W.~K. Wootters and D.~M. Sussman, ``Discrete phase space and
  minimum-uncertainty states,'' {\em arXiv:0704.1277}, 2007.

\bibitem{GAP4}
The GAP~Group, {\em {GAP -- Groups, Algorithms, and Programming, Version
  4.10.2}}, 2019.

\bibitem{CAP}
S.~Gutsche, S.~Posur, and {\O}.~Skarts{\ae}terhagen, ``On the syntax and
  semantics of $\mathtt{CAP}$,'' in {\em In: O. Hasan, M. Pfeiffer, G. D. Reis
  (eds.): Proceedings of the Workshop Computer Algebra in the Age of Types,
  Hagenberg, Austria, 17-Aug-2018}, published at
  \url{http://ceur-ws.org/Vol-2307/}, 2018.

\bibitem{GH96}
M.~Geck, G.~Hiss, F.~L{\"u}beck, G.~Malle, and G.~Pfeiffer, ``{\sf CHEVIE} --
  {A} system for computing and processing generic character tables for finite
  groups of {L}ie type, {W}eyl groups and {H}ecke algebras,'' {\em Appl.
  Algebra Engrg. Comm. Comput.}, vol.~7, pp.~175--210, 1996.

\bibitem{GAP3}
M.~S. et~al., {\em {GAP} -- {Groups}, {Algorithms}, and {Programming} --
  version 3 release 4 patchlevel 4}.
\newblock RWTHLDFM, RWTH-A, 1997.

\end{thebibliography}
\bibliographystyle{ieeetr}

\appendix

\section{Construction of uniform and rigidly symmetric measurement assemblages}
\label{app:construction}

In this section we describe the algorithm to construct measurement assemblages from a selected symmetry group.
We start with a subsection summarising the basic notions of group action and group representation.
Readers who are familiar with these concepts can skip this subsection.

\subsection{Groups, group action and group representation}
By a group, we always consider an abstract set $G$ with a multiplication defined such that
\begin{itemize}
  \item[(G1)] for all $g_1,g_2,g_3 \in G$, $g_1 (g_2 g_3) = (g_1 g_2) g_3$;
  \item[(G2)] there is $1 \in G$ such that for all $g \in G$, $g 1 = 1 g = g$;
  \item[(G3)] for all $g \in G$, there is $g^{-1} \in G$ such that $gg^{-1} = g^{-1}g =1$.
\end{itemize}
It this work, we maintain the viewpoint that a group is defined in this abstract sense, rather than a concrete realisation of group as permutations or matrices, which arises as the group acts on a set or a vector space.

An abstract group can act on different sets of different natures.
More precisely, let $S$ be a finite or infinite set, a group action is a map $\varphi: G \to \mathcal{F}(S)$, where $\mathcal{F} (S)$ is the set of invertible maps on $S$, such that
\begin{equation}
  \varphi(g_1 g_2) = \varphi(g_1) \varphi(g_2)
\end{equation}
for all $g_1, g_2 \in G$.
For $S$ being finite, $\mathcal{F} (S)$ is simply the group of permutations that permute elements of $S$.
When $S$ has more algebraic structure (such as a vector space), $\mathcal{F} (S)$ may be limited to maps that conserve the corresponding algebraic structure (such as linear transformations).
In particular, if $S$ is a Hilbert space, and $\mathcal{F} (S)$ contains the unitary transformations, then the group action is said to be a unitary representation of the group $G$.
In this case, the action is often denoted by $U$, and $U_g$ denotes the unitary operator corresponding to element $g$.

In practice, the action of a group $G$ on a set $S$ can be thought of as the mathematical description of the symmetry of $S$ via the group $G$.
For $g \in G$, $\varphi(g)$ is a map from $S$ to $S$.
As a general convention that has been used in the main text, for $x \in S$, the element $\varphi(g) [x] \in S$ is often simply denoted as $g(x)$.
This convention is applied throughout, except for unitary representations.

For $x \in S$, the set $G(x) =\{g(x): g \in G\}$ is called the orbit of $x$.
Under the action of $G$, $S$ is partitioned into different orbits.
The action is said to be transitive if $S$ contains a single orbit.
The construction of a highly symmetric measurement assemblage is in fact the construction of an orbit of $G$ with certain particular requirements.
We therefore are interested in the classification of orbits of $G$.

The orbits of $G$ can be characterised via the concept of stabiliser (sub)groups.
More precisely, let $G$ act on $S$.
For $x \in S$, $G_x=\{g \in G: g(x) =x\}$ is called the stabiliser group (or the isotropy group) of $x$.
It is straightforward to show that the number of elements in the orbit of $x$ can be given by $\abs{G(x)} = G/\abs{G_x}$ (note that the size of any subgroup of $G$ divides the size of $G$).
Moreover, if $x$ and $y$ are in the same orbit, the stabiliser groups $G_x$ and $G_y$ are conjugated, i.e., $G_x=g G_y g^{-1}$ for some $g \in G$.
In fact, two orbits are said to be of the same \emph{type} if the stabiliser groups of the elements in the orbits are conjugated.
Therefore classification of orbits of $G$ according to their types is the same as classification of conjugacy classes of subgroups of $G$.

The above concepts are sufficient to support our further discussions.
Readers who are interested in more details are referred to Ref.~\cite{Armstrong2010a,Serre1977a}.

\subsection{Ideas of the construction}

Starting with a group $G$ and a unitary representation $U: G \to \U(d)$, we would like to construct a family of uniform and rigidly symmetric projective measurement assemblages.

In our example, $G$ is a complex reflection group, which is a matrix group.
The representation is simply the natural action of the matrices on the vector space where the group is defined (with an appropriate inner product).
The group acts on the space of matrices by means of conjugation.

The first step in the construction is to construct an orbit of $G$.
With a generating projection $P$ at hand the orbit is given by $\{U_g P U_g^{-1}: g \in G\}$.
(A note regarding the terminology: in this paper, projectors and projections are considered as synonyms.)
Such a projection $P$ can be identified by its stabiliser group.
Moreover, by the rigidity requirement, the stabiliser group is required to commute with exactly two proper projections.
In the language of linear representation theory, this implies that the representation $U$ restricted to the stabiliser group has exactly two irreducible subrepresentations; a fact that can be checked easily via character theory~\cite{Serre1977a}.

Therefore, we can start by enumerating all conjugacy classes of subgroups of $G$ and filter those that have exactly two irreducible subrepresentations.
Choosing the projection onto one of these, we can generate its orbit under the action of $G$.
In this orbit, subsets of projections are grouped to form projective measurements if they sum up to the identity operator.
For construction of nonprojective measurements (i.e., positive-operator valued measures -- POVMs), we only require that the sum of the subsets is proportional to the identity operator.
The last step is to check and exclude the orbits that do not fulfil the covariance condition~\eqref{eq:covariance} in the main text.

\subsection{The construction algorithm}

This algorithm summarises the above discussion.
The starting point is a group $G$ and a unitary representation $U: G \to \U(d)$.
\begin{enumerate}
  \item Enumerate all conjugacy classes of subgroups of $G$.
    Each of the conjugacy classes will be a candidate for the stabiliser group at a point.
  \item Find all classes whose representatives have exactly two subrepresentations.
    As the representatives are the stabiliser groups, this ensures the rigidity of the assemblage.
  \item For each of such classes, take the projection onto one of the irreducible representations.
    Find the stabiliser group as $G$ acts on it by conjugation.
    As a matter of fact, the stabiliser group can be bigger than the corresponding original representative of the conjugacy class of subgroups.
    Reclassify all the obtained projections according to their stabiliser groups.
  \item For each generating projection, obtain its orbit as $G$ acts on it via conjugation and group the projections in the orbit into orthogonal subsets.
    \label{enum:group}
  \item Test if the group action preserves these orthogonal subsets, i.e., respects the covariance condition~\eqref{eq:covariance}.
\end{enumerate}

Minor adaptation is sufficient to construct also nonprojective measurements.
To this end, we choose a number of outcomes $n$ and look for a combination of projections whose sum is proportional to the identity.
This simple procedure turns out to be ultimately related to the notion of tight frames~\cite{Wal18}.
See Table~\ref{tab:povm} for the different nonprojective measurement assemblages that we found by means of the complex reflection groups~\cite{ST54}.

Note that in dimension $d=2$ we recover all the projective measurement assemblages defined by the regular polytope (platonic solids).
Also for nonprojective measurements in dimension $d=2$, we recover the known interesting structures such as the regular polyhedron compound discussed in~\cite{CGGZ19}.

\begin{table*}[ht!]
  \begin{tabular}{|c|c|c|c|p{4cm}|c|c|}
    \hline
    $~d~$                & ~Group~                  & $~n~$                & $\abs{M}$            & \centering Comments                                                  & $\quad\alpha^\ast=\max\{\eta:A^\eta\ \mathrm{compatible}\}\quad$ & $\quad\beta^\ast=\max\{\eta:\bar{A}^\eta\ \mathrm{compatible}\}\quad$ \\ \hline
    \multirow{8.7}{*}{2} & \multirow{3.7}{*}{ST 8}  & 3                    & 4                    & \centering Cuboctahedron                                             & \multicolumn{2}{c|}{$\frac{1}{\sqrt{2}}\approx0.7071$}                                                                                   \\ \cline{3-7}
                         &                          & \multirow{1.7}{*}{4} & \multirow{1.7}{*}{2} & \centering $\qquad$$\qquad$Cube$\qquad$$\qquad$ Tetrahedron compound & \multicolumn{2}{c|}{\multirow{1.7}{*}{$\sqrt{\frac23}\approx0.8165$}}                                                                    \\ \cline{3-7}
                         &                          & 4                    & 3                    & \centering Cuboctahedron                                             & \multicolumn{2}{c|}{$\sqrt{\frac23}\approx0.8165$}                                                                                       \\ \cline{2-7}
                         & \multirow{5.4}{*}{ST 16} & 3                    & 10                   & \centering Icosidodecahedron                                         & \multicolumn{2}{c|}{$\sqrt{\frac{5+2\sqrt{5}}{20}}\approx0.6882$}                                                                        \\ \cline{3-7}
                         &                          & \multirow{1.7}{*}{4} & \multirow{1.7}{*}{5} & \centering $\qquad$Dodecahedron$\qquad$ Tetrahedron compound         & \multicolumn{2}{c|}{\multirow{1.7}{*}{$\sqrt{\frac{5+2\sqrt{5}}{15}}\approx0.7947$}}                                                     \\ \cline{3-7}
                         &                          & 5                    & 6                    & \centering Icosidodecahedron                                         & \multicolumn{2}{c|}{$\sqrt{\frac{7+3\sqrt{5}}{24}}\approx0.7558$}                                                                        \\ \cline{3-7}
                         &                          & \multirow{1.7}{*}{6} & \multirow{1.7}{*}{5} & \centering $\qquad$Icosidodecahedron$\qquad$ Octahedron compound     & \multicolumn{2}{c|}{\multirow{1.7}{*}{$\sqrt{\frac{5+\sqrt{5}}{10}}\approx0.8507$}}                                                      \\ \hline
    \multirow{4}{*}{3}   & ST 24                    & 4                    & 7                    &                                                                      & $\approx0.5349$                                                  & $\approx0.9190$                                                       \\ \cline{2-7}
                         & \multirow{3}{*}{ST 27}   & 4                    & 15                   &                                                                      & $\approx0.5193$                                                  & $\approx0.7643$                                                       \\ \cline{3-7}
                         &                          & 6                    & 6                    &                                                                      & $\frac{5+3\sqrt{5}+\sqrt{790+270\sqrt{5}}}{80}\approx0.6130$     & $\frac{1+\sqrt{5}+\sqrt{30-6\sqrt{5}}}{8}\approx0.9135$               \\ \cline{3-7}
                         &                          & 6                    & 10                   &                                                                      & $\gtrapprox0.5973$                                               & $\frac{2+3\sqrt{5}}{10}\approx0.8708$                                 \\ \hline
    \multirow{3}{*}{4}   & ST 29                    & 5                    & 16                   &                                                                      & $\gtrapprox0.4164$                                               & $\approx0.8954$                                                       \\ \cline{2-7}
                         & ST 30                    & 5                    & 60                   &                                                                      & $\geq\frac{20+7\sqrt5+\sqrt{2115+910\sqrt5}}{180}\approx0.5560$  & $\gtrapprox0.9163$                                                    \\ \cline{2-7}
                         & ST 31                    & 5                    & 96                   &                                                                      & $\geq\frac{14+\sqrt{679}}{96}\approx0.4173$                      & $\gtrapprox0.8011$                                                    \\ \hline
  \end{tabular}
  \caption{
    Nonprojective rank-one measurement assemblages constructed from the complex reflection groups and their incompatibility properties.
    The number $d$ is the dimension, the groups are given through their Shephard--Todd (ST) number~\cite{ST54}, $n$ is the number of outcomes, and $\abs{M}$ is the number of measurements.
    The last two columns illustrate the power of symmetry by giving analytically two interesting incompatibility properties, $\alpha^\ast$ and $\beta^\ast$ as defined in the main text.
    The values can all be exactly represented (with radicals), but large representations are converted into numerical values.
    Equivalently, these quantities correspond to the noise threshold for steering the isotropic (left) and the Werner (right) states (see Appendix~\ref{app:interpretation} for details).
    For too large $\abs{M}$, only bounds on $\alpha^\ast$ and $\beta^\ast$ can be obtained, thanks to a heuristic method inspired by statistical mechanics (see Appendix~\ref{app:heuristic}).
  }
  \label{tab:povm}
\end{table*}

\section{Simplification of the computation of incompatibility by symmetry}
\label{app:simplification}

In this appendix, we demonstrate how to simplify computations involving symmetric measurement assemblages.
We sketch the general principle of using symmetry in convex optimisation problems.
This is followed by an illustration on the problem of computing the incompatibility robustness with respect to white noise \cite{DSFB19,DFK19}.
Other related problems are later discussed.

As in the main text, $A$ denotes a measurement assemblage defined on the bundle of outcomes $(\Omega,\pi,M)$.
The symmetry group of the assemblage is described by a group $G$ acting on $(\Omega,\pi,M)$ together with a unitary representation $U$ of $G$ on $\CC^d$.
In addition, $M_d(\CC)$ denotes the space of matrices of size $d$ with elements in $\CC$, $M_d^H(\CC)$ its subspace of hermitian matrices, and $M_d^+(\CC)$ its positive cone.

\subsection{Symmetry of a convex optimisation problem}
\label{app:simplification_general}

In the most general form, we consider the problem of minimising a symmetric convex function over a symmetric domain.
More specifically, let $\mathcal{X}$ be a real vector space and $D$ be a convex subset of $\mathcal{X}$.
We are concerned with the following problem,
\begin{equation}
  \gamma^{\ast} = \min_{x} f(x) \mbox{ for $x \in D$},
  \label{eq:optimisation_general}
\end{equation}
where the objective function $f:\mathcal{X} \to \RR$ is assumed to be convex.

The symmetry of the problem is described by a linear action of a group $G$ on $\X$ such that both $f$ and $D$ are invariant under $G$, that is,
\begin{equation}
  \left\{\begin{array}{l}f[g(x)]=f(x)\\ g (x) \in D \end{array}\right.\mbox{ for all $x \in D$ and $g \in G$}.
\end{equation}
The standard argument from group theory says that, from an optimal solution $x^\ast$, one can construct another one that is fixed under $G$, namely, $\sum_{g\in G}g (x^\ast)/|G|$.
Thus the optimisation can be performed only on the smaller set of variables invariant under $G$, namely,
\begin{equation}
  \gamma^{\ast} = \min_{x} f(x) \mbox{ for $x \in D^\ast$},
\end{equation}
where $D^\ast=\{x\in D:g(x)=x\quad\forall g\in G\}$.
In particular, if there is a unique point in $D$ invariant under $G$, it must be the optimal solution.

In practice, the domain $D$ is often also generated by a family of invariant functions and symmetry also allows one to reduce the number of constraints.
The formal description of this procedure is somewhat cumbersome, and we will discuss it directly in the concrete situations.

\subsection{The primal problem}
\label{app:primal}

The computation of the incompatibility robustness is of the form
\begin{align}
  \alpha^\ast = \max_{\eta,F} \quad &  \eta \label{eq:primalapp} \\
  \mbox{s.t.} \quad & \!\!\!\!\sum_{s \in \Gamma(\Omega)} \delta_{s[\pi(z)],z} F_s = A_{z}^\eta\mbox{ for all $z \in \Omega$} \nonumber \\
  & F_s \ge 0 \mbox{ for all $s \in \Gamma(\Omega)$},\nonumber
\end{align}
where $A^\eta_z=\eta A_z + (1-\eta) \Tr (A_z) \I/d$.

The problem~\eqref{eq:primalapp} is a special case of~\eqref{eq:optimisation_general}.
In this case the vector space $\mathcal{X}$ is the space of points $x=(\eta,F)$, where $\eta \in \RR$ and $F:\Gamma(\Omega) \to M_d^H(\CC)$.
It is clear that the objective function and the domain are convex.

The crucial observation is that as long as the bundle of outcomes $(\Omega,\pi,M)$ has a certain symmetry described by a group $G$, then the space of sections $\Gamma(\Omega)$ inherits a symmetry with the same group $G$.
Specifically, the action of $G$ on the set of sections $\Gamma(\Omega)$ is defined by
\begin{equation}
  [g(s)](x) = g (s [g^{-1} (x)]) \mbox{ for all $g \in G$ and $x\in M$}.
  \label{eq:action-sections}
\end{equation}

Let us now study the action of $G$ on the space of the variable $(\eta,F)$.
The group $G$ leaves $\eta$ fixed, and transforms $F$ by
\begin{equation}
  [g(F)]_s=  U_g F_{g^{-1}(s)} U^{-1}_g.
\end{equation}
The objective function $f(x)=\eta$ and the domain of the problem are easily seen to be invariant under $G$.
According to our general remark in Section~\ref{app:simplification_general}, this implies that we can assume $g(F)=F$, or, written explicitly,
\begin{equation}
  F_s=U_g F_{g^{-1}(s)} U^{-1}_g\mbox{ for all $g \in G$ and $s\in\Gamma(\Omega)$}.
\end{equation}
This says that the values of $F$ at sections that are related by a symmetry element $g$ are related by the corresponding unitary $U_g$.
As a matter of fact, the number of variables of the optimisation problem can be reduced to the number of equivalence classes of $\Omega$ under the action of $G$.
(Here and in the following, the number of variables refers to the number of matrices in the SDP, that is, $\abs{\{F_s: s \in \Gamma(\Omega) \}}=\abs{\Gamma(\Omega)}$.)
In a similar way, the number of constraints can also be reduced.
Indeed, for a symmetric $F$ (that is, $g(F)=F$), if two outcomes $z_1$ and $z_2$ are equivalent under the action of $G$, then $\sum_{s \in \Gamma(\Omega)} \delta_{s[\pi(z_1)],z_1} F_s = A_{z_1}^\eta$ implies $\sum_{s \in \Gamma(\Omega)} \delta_{s[\pi(z_2)],z_2} F_s = A_{z_2}^\eta$.
(Here and in the following, the number of constraints refers to the number of matrix equalities/inequalities in the SDP.)

For example, for the dodecahedron assemblage, the original problem with $2^{10}$ variables and $10$ constraints reduces to one with $20$ variables and $1$ constraint.
But as we mentioned in the main text, this is \emph{not} the whole story; the symmetry has much deeper implications when we approach the problem from the dual perspective.

\subsection{The dual problem}

The dual of the problem~\eqref{eq:primalapp} is written as~\cite{DSFB19}
\begin{align}
  \label{eq:dualapp}
  \alpha^\ast =\min_X \quad & 1 + \sum_{z \in \Omega} \Tr(X_z A_z) \\
  \mbox{s.t.} \quad & 1 + \sum_{z \in \Omega} \Tr(X_z A_z) \ge \frac{1}{d} \sum_{z \in \Omega} \Tr (A_z) \Tr (X_z) \nonumber \\
  & \sum_{z \in \Omega} \delta_{s[\pi(z)],z} X_z \ge 0 \qquad \forall s \in \Gamma(\Omega).\nonumber
\end{align}

Now the variable of the optimisation problem is $X: \Omega \to M_d^H(\CC)$ and $G$ acts by $[g( X)]_z=U_g X_{g^{-1}(z)} U_g^{-1}$.
It is again easy to see that given $A$ symmetric, both the objective function and the domain are symmetric.
This implies that one can impose the symmetry constraint on the variable of the problem, that is, $g (X)=X$, or
\begin{equation}
  X_z = U_g X_{g^{-1}(z)} U_g^{-1}\mbox{ for all $g \in G$ and $z\in\Omega$}.
  \label{eq:symmetry-X}
\end{equation}
Thus $X$ has the same symmetry as $A$.
Also, again, once the symmetry is imposed on the variable, the number of constraints can also be reduced.

Let $\bar{\Omega}$ denote a set of representatives of equivalence classes of $\Omega$ and $\bar{\Gamma}(\Omega)$ denote a set of representatives of equivalence classes of $\Gamma(\Omega)$.
One has the following decomposition
\begin{align}
  \sum_{z\in\Omega} \Tr(X_zA_z) &= \sum_{z_i\in\bar{\Omega}}\sum_{z\in [z_i]}\Tr(X_zA_z)\nonumber\\
  &= \sum_{z_i\in\bar{\Omega}}\frac{1}{\abs{G_{z_i}}}\sum_{g\in G}\Tr(X_{g (z_i)} A_{g (z_i)})\nonumber\\
  &= \sum_{z_i\in\bar{\Omega}}\frac{\abs{G}}{\abs{G_{z_i}}} \Tr(X_{z_i} A_{z_i}).\nonumber
\end{align}
A similar manipulation can be performed on the second constraint so that Eq.~\eqref{eq:dualapp} can eventually be computed through the simplified form given in Eq.~\eqref{eq:dual-sym} below.
More importantly, this symmetrised SDP in fact did not implement yet the full symmetry in the variable $X$ in Eq.~\eqref{eq:symmetry-X}.
In addition to the present constraints, one can require from Eq.~\eqref{eq:symmetry-X} that $X_{z_i}$ commutes with all of $U(G_{z_i})$.
This in fact can significantly simplify the problem, as it implies that $X_{z_i}$ must have a certain block structure dictated by the irreducible decomposition of $U(G_{z_i})$~\cite{Serre1977a}.
The case of uniform and rigidly symmetric assemblages discussed in the main text is an example where this constraint implies that $X$ has only two free parameters.
Below we extend the details of this discussion.
\begin{widetext}
  \begin{align}\label{eq:dual-sym}
    \alpha^\ast = \min_{\{X_{z_i}\}} \quad & 1 +  \sum_{z_i \in \bar{\Omega}} \frac{\abs{G}}{\abs{G_{z_i}}} \Tr(X_{z_i} A_{z_i}) \\
    \mbox{s.t.} \quad  & 1 +  \sum_{z_i \in \bar{\Omega}} \frac{\abs{G}}{\abs{G_{z_i}}} \Tr(X_{z_i} A_{z_i}) \ge \frac{1}{d}  \sum_{z_i \in \bar{\Omega}} \frac{\abs{G}}{\abs{G_{z_i}}} \Tr(X_{z_i})\Tr(A_{z_i})\nonumber\\
    & \!\sum_{z_i \in \bar{\Omega}} \frac{1}{\abs{G_{z_i}}} \sum_{g \in G} \delta_{[g^{-1}( s_j)][\pi(z_i)],z_i} U_g X_{z_i} U_{g}^{-1} \ge 0\qquad\forall s_j \in \bar{\Gamma}(\Omega) \nonumber.
  \end{align}
\end{widetext}

\section{Incompatibility of uniform, rigidly symmetric assemblages}
\label{app:rigid-symmetry}

In the main text, we have shown that for uniform and rigidly symmetric assemblages the condition~\eqref{eq:symmetry-X} on the dual variable $X$ implies that $X$ has a rather specific form, namely, 
\begin{equation}
  X_z = a \I + b A_z,
  \label{eq:ansatz}
\end{equation}
for some $a$ and $b$ that we are now going to fix.

\subsection{Strategy to fix the parameters}
\label{app:parameters}

For the solution~\eqref{eq:ansatz} to satisfy the constraints of Eq.~\eqref{eq:dualapp}, we need
\begin{align}
  1 &+\Tr \left( \sum_{z\in\Omega} X_zA_z \right) - \frac{1}{d}\sum_{z\in\Omega}\Tr (A_z)\Tr (X_z) \nonumber \\
  &= 1+ b \sum_{z\in\Omega}\left[\Tr (A_z^2)-\frac{1}{d}\left(\Tr A_z\right)^2\right] \geq 0,
  \label{eqn:firstcond}
\end{align}
and
\begin{align}
  \sum_{z\in\Omega}\delta_{s[\pi(z)],z}X_z &= \sum_{z\in\Omega}\delta_{s[\pi(z)],z} \left(a\I+bA_z\right) \nonumber \\
  &= a\abs{M}\I +b \sum_{x\in M} A_{s(x)} \geq 0.
  \label{eqn:secondcond}
\end{align}
First, Eq.~\eqref{eqn:firstcond} is saturated when we pick $b=-1/Z$ with
\begin{equation}
  Z=\sum_{z\in\Omega}\Tr A_z^2-\frac{d\abs{M}^2}{\abs{\Omega}},
\end{equation}
where we have used the uniformity to get $\Tr (A_z)=d\abs{M}/\abs{\Omega}$.
Second, Eq.~\eqref{eqn:secondcond} is saturated when we pick $a=-{b\lambda}/{\abs{M}}=\lambda/(\abs{M}Z)$ with
\begin{equation}
  \lambda = \max_{s\in\Gamma(\Omega)} \left\| \sum_{x\in M} A_{s(x)}\right\|_\infty\!\! ,
  \label{eq:lambda}
\end{equation}
which is the largest eigenvalue of all the operators $\sum_{x\in M} A_{s(x)}$ for $s\in\Gamma(\Omega)$.
Note that the computation of $\lambda$ requires an optimisation over all sections.
This in the worst case can be done by enumerating all the sections.
With these values for $a$ and $b$, Eq.~\eqref{eq:ansatz} becomes
\begin{equation}
  X_z=\frac1Z\left(\frac{\lambda}{\abs{M}}\I-A_z\right),
  \label{eqn:feasible}
\end{equation}
which is an optimal point for the problem~\eqref{eq:dual}.
Thus we finally get
\begin{equation}
  \alpha^\ast= \frac{d}{Z}\left(\lambda-\frac{\abs{M}^2}{\abs{\Omega}}\right).
  \label{eq:tight-bound}
\end{equation}
For rank-one projective measurements we have $Z=(d-1)\abs{M}$ and $\abs{\Omega}=d\abs{M}$ so that this simplifies to
\begin{equation}
  \alpha^\ast=\frac{\lambda-\frac{\abs{M}}{d}}{\abs{M}-\frac{\abs{M}}{d}}.
\end{equation}
Note that, though these formulae look like the bounds obtained in Ref.~\cite{DSFB19}, they are of a complete different nature as they are here guaranteed to be equalities thanks to the symmetry.

Similarly, one gets
\begin{equation}
  \beta^\ast= \frac{d(d-1)}{Z}\left(\frac{\abs{M}^2}{\abs{\Omega}}-\mu\right),
  \label{eq:tight-bound-beta}
\end{equation}
where
\begin{equation}
  \mu=\!\!\min\limits_{s\in\Gamma(\Omega)} \left\| \sum_{x\in M} A_{s(x)}\right\|_\infty.
\end{equation}
For rank-one projective measurement assemblages, one finds
\begin{equation}
  \label{eq:beta-proj}
  \beta^\ast=1-\frac{\mu}{\abs{M}}d.
\end{equation}

\subsection{Mapping to statistical mechanics}
\label{app:statmech}

Curiously, the problem of computing $\lambda$ in Eq.~\eqref{eq:lambda} can be mapped to a statistical mechanics model.
While this mapping does not solve the computational problem, it brings some interesting insight and suggests a heuristic approach to the problem (see Sec.~\ref{app:heuristic}).

To this end, consider a system of $\abs{M}$ so-called Potts spins (each corresponding to a measurement) coupled to a continuous variable $\psi$ on the general qudit Bloch sphere of dimension $d$ (i.e., the set of pure states).
A Potts spin (corresponding to a measurement) is simply a classical system of a finite number of states (here the number of states is simply the number of outcomes of each measurement)~\cite{Mezard2009a}.
Let us emphasise that $\psi$ is a \emph{classical} random variable whose values are points on the Bloch sphere.
The state of the whole system is specified by a section of outcomes (states of all Potts spins) $s$ and the value of $\psi$.
Consider the Hamiltonian
\begin{equation}
  H(s,\psi)= -\sum_{x \in M} \bra{\psi} A_{s(x)} \ket{\psi},
  \label{eq:hamiltonian}
\end{equation}
whose ground state energy is
\begin{equation}
  \min_{s,\psi} {H(s,\psi)}=-\lambda.
\end{equation}
Statistical mechanics suggests to look at the system at finite temperature $T$, where it follows the Boltzmann distribution
\begin{equation}
  p(s,\psi)= \frac{1}{Z} e^{-H(s,\psi)/T},
\end{equation}
where $Z$ is the partition function~\cite{Mezard2009a}
\begin{equation}
  Z= \sum_{s \in \Gamma(\Omega)} \int \d \omega (\psi) \sum_{x \in M} e^{-H(s,\psi)/T},
\end{equation}
with $\omega$ being the Haar measure over the qudit Bloch sphere.

\begin{figure}[h]
  \centering
  \includegraphics[width=0.40\textwidth]{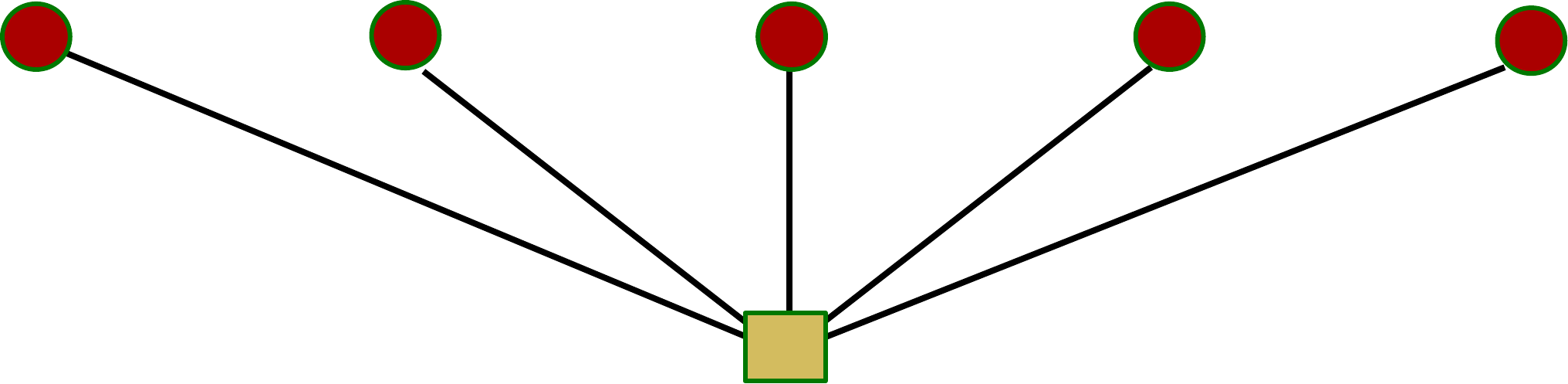}
  \caption{The tree representing the interaction between the continuous variable $\psi$ (square) with Potts spins (circle).}
  \label{fig:3}
\end{figure}

Let us have a closer look at the Hamiltonian~\eqref{eq:hamiltonian}.
Crucially, it is a sum of pairwise interactions since each term involves only two variables, $\psi$ and a Potts spin.
These pairwise interactions form a tree, i.e., a graph without any loop.
One can present this Boltzmann distribution by a graph~\cite{Mezard2009a} as in Figure~\ref{fig:3}.

Being a model defined on a tree implies that the method of message passing (also known as transfer matrix) to find the marginal distributions for each Potts spin is exact~\cite{Mezard2009a}.
As the marginal distribution for each Potts spin carries the information of the ground state, $\lambda$ can also be computed.
Unfortunately, in order to perform this algorithm, integration over the Bloch sphere is required, which is difficult to carry out particularly when one works at zero temperature.
One option is to approximate the Bloch sphere by means of a complex projective design and to carry out the algorithm at finite temperature.
The result can then be extrapolated to zero temperature.
Another option is to use simulated annealing to gradually cool the system down to its ground state~\cite{Press1992a}.
These methods can be considered when a good approximation for the ground state for a large system is required.
However, in this work, we use an even simpler argument to approximate the ground state energy.

By direct investigation in many cases, one can heuristically expect that the system presents \emph{no frustration} and is of \emph{ferromagnetic type}~\cite{Mezard2009a,Dotsenko1994a}.
This means that at low temperature, the Potts spins behave collectively in a way that they prefer to `align with each other.'
The number of ground states of the system simply equals the number of states of one Potts spin (if no other degeneracy is present).
Moreover, they all have the same energy.
This is in strong contrast with systems with antiferromagnetic interactions and involving loops~\cite{Dotsenko1994a}.
In these cases, the energy can have a very complicated energy landscape with many local minima~\cite{Dotsenko1994a}.
As the system is cooled down, it can easily be trapped into a local minimum, and tunnelling between local minima can slowly happen, a behaviour reflecting glassy phase transitions and ageing~\cite{Dotsenko1994a}.

Based on the heuristic assumption that the system is of ferromagnetic type, we can expect the following algorithm to give some good approximation to the ground state energy.
One starts by fixing the state of one of the Potts spin in order to break the symmetry between the different ground states.
This forms a seed for a phase transition to happen~\cite{Goldenfeld1972a}.
Next, we seek the next Potts spin, which is chosen such that the new droplet system of two Potts spins has the lowest possible energy (i.e., we look for the nearest neighbour of the original one).
While the state of the old spin is fixed, the state of the added spin is chosen so that the new energy is minimised.
One then continuously add more spins until the whole system is exhausted.
The state obtained is then expected to be close to the ground state of the system.

\subsection{Pseudocode to heuristically estimate \texorpdfstring{$\lambda$}{lambda}}
\label{app:heuristic}

For the convenience of readers who are unfamiliar with statistical mechanics models, we write here the pseudocode without referring to the above underlying idea.

The purpose is to find a section $s$ such that $\lambda$ in Eq.~\eqref{eq:lambda} is minimised.
\begin{enumerate}
  \item Start with a measurement $x$ and an outcome $s(x)$ (arbitrarily because of the uniformity of the assemblage).
    Define $S=A_{s(x)}$ and $\lambda=\|S\|_\infty$, which is then $1$ for projective measurements.
  \item\label{enum:algo} Search for the measurement $y$ and outcome $s(y)$ such that $\lambda=\|S+A_{s(y)}\|_\infty$ is the biggest among all possibilities, where $S$ was defined in the previous step.
    Then define the new value of $S$ to be $S+A_{s(y)}$.
  \item Repeat step \ref{enum:algo} until all measurements are selected.
\end{enumerate}
In step \ref{enum:algo}, degeneracies can occur, i.e., many candidates for $y$ can be found.
In this case, we arbitrarily select one of them.

When comparing with the exact values obtained by enumeration (when possible), we find that the procedure almost always gives the optimal sections.
However, there are a few exceptions: for instance, MUBs in dimension eight or the measurement assemblage with 20 projective measurements in dimension three obtained with ST~27.
In fact, we expect that future research will be able to pinpoint the condition under which this procedure gives the exact maximal $\lambda$; the analogy with statistical physics could give important hints on this.

\section{Consequence for Einstein--Podolsky--Rosen steering}
\label{app:steering}

Measurement incompatibility has direct consequences for the so-called Einstein--Podolsky--Rosen (EPR) steering~\cite{Wiseman2007a}.
Here we give a brief introduction to EPR steering and explain the connection, so that our results can be directly interpreted in this context.

\subsection{Introduction to EPR steering}

\begin{figure}[h]
  \centering
  \includegraphics[width=0.4\textwidth]{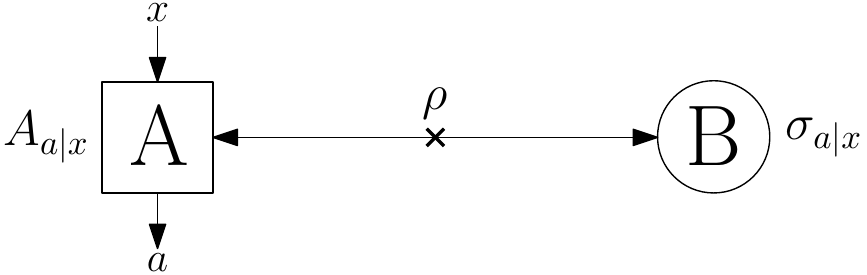}
  \caption{Steering scenario, where Alice makes measurements in assemblages $A$, steering Bob's system to the corresponding conditional states.}
  \label{fig:4}
\end{figure}

EPR steering is an intermediate scenario that lies in between entanglement and nonlocality~\cite{Wiseman2007a}.
It involves two parties, usually referred to as Alice and Bob, who share a bipartite quantum state $\rho$.
Applying a measurement assemblage on her side, Alice produces a state assemblage on Bob's side (see Fig.~\ref{fig:4}).
This state assemblage is said to be steerable if no so-called local hidden state model can explain the statistics he collects~\cite{Wiseman2007a}.

More formally, the state assemblage created on Bob's side corresponding to Alice's measurement assemblage $A$ defined on the bundle of outcome $(\Omega,\pi,M)$ is $\tau: \Omega \to M_d^+(\CC)$, $\tau_z=\Tr_A[(A_z\otimes\I)\rho]$.
Here $\Tr_A$ denotes the partial trace over Alice's system.
To make the connection to the familiar notation~\cite{Uola2019a}, recall that in the discrete bundle formalism introduced in the main text, $z=(a|x)$.
This state assemblage is unsteerable when a local hidden state model can explain it, specifically, when there exists $\sigma: \Gamma (\Omega) \to M_d^+(\CC)$ such that
\begin{equation}
  \label{eqn:decomp}
  \tau_z= \sum_{s \in \Gamma (\Omega)} \delta_{s[\pi (z)],z} \sigma_s.
\end{equation}
Again, it is easy to identify this with the definition of local hidden state model in the more familiar notation such as in, e.g., Refs~\cite{CS16b,Uola2019a}.

\subsection{Interpretation of the incompatibility robustness}
\label{app:interpretation}

Observing Eq.~\eqref{eqn:decomp}, one may already anticipate the intimate (mathematical) connection between EPR steering and measurement incompatibility.
Indeed, it is well-known that finding a parent measurement for a measurement assemblage (see Eq.~\eqref{eq:compatibility-def} in the main text) is the same task as finding a local hidden state model for a state assemblage~\cite{QVB14,UBGP15}.

Consider an isotropic state defined on a bipartite system of dimension $d \times d$ as
\begin{equation}
  \rho_\mathrm{iso}^\zeta = \zeta \ketbra{\Phi^+}{\Phi^+} + (1-\zeta) \frac{\I}{d} \otimes \frac{\I}{d},
\end{equation}
where $\ket{\Phi^+}= \sum_{k=1}^{d} \ket{k,k}/\sqrt{d}$.
Here $\zeta\in[0,1]$ is also referred to as a noise.
Performing a measurement assemblage $A$ on Alice's side produces a state assemblage on Bob's side as
\begin{align}
  \tau_z^\zeta&= \Tr_A [(A_z\otimes\I)\rho_\mathrm{iso}^\zeta]\\
  &=\frac1d\left(\zeta A_z^T+(1-\zeta)\Tr (A_z)\frac{\I}{d}\right),
\end{align}
which is simply a rescaling of the transposition of $A^\zeta$ defined in the main text.
Thus the noise threshold from which the isotropic state can be steered by using the measurements in $A$ is precisely $\alpha^\ast$.

Likewise, consider the Werner state~\cite{Werner1989a} defined by
\begin{equation}
  \rho_W^\zeta=\zeta\frac{2P_d^{(-)}}{d(d-1)}+(1-\zeta)\frac{\I}{d}\otimes\frac{\I}{d},
\end{equation}
where $P_d^{(-)}$ is the projection onto the antisymmetric subspace of $\CC^d\otimes\CC^d$.
Performing a measurement assemblage $A$ on Alice's side produces a state assemblage on Bob's side as
\begin{align}
  \tau_z^\zeta&=\Tr[(A_z\otimes\I)\rho_W^\zeta]\\
  &=\frac1d\left(\zeta\frac{\Tr (A_z) \I-A_z}{d-1}+(1-\zeta)\Tr (A_z)\frac{\I}{d}\right),
\end{align}
which is a rescaling of $\bar{A}^\zeta$ defined in the main text.
Thus the noise threshold from which the Werner state can be steered by using the measurements in $A$ is precisely $\beta^\ast$.

\subsection{Finite measurement assemblages that are more incompatible than all dichotomic measurements}

As we mentioned in the main text there are three measurement assemblages in Table~\ref{tab:marvel} that are more incompatible than all dichotomic measurements.
Thanks to the above connection, this immediately implies that they can exploit steering with a Werner state while all dichotomic measurements fail to do so.
The existence of this was nonconstructively established in Ref.~\cite{Nguyen2019b}; here we give an explicit construction.

For the sake of completeness, let us recall the threshold obtained therein.
For the isotropic state $\rho_\mathrm{iso}^\zeta$ in dimension $d$, the set of all two-outcome measurements can demonstrate steering for
\begin{equation}
  \zeta\geq1-d^{-1/(d-1)}.
  \label{eq:dicho-iso}
\end{equation}
This approximately evaluates to $0.4226$ for $d=3$ and $0.3700$ for $d=4$.
For the Werner state $\rho_W^\zeta$ in dimension $d$, the set of all two-outcome measurements can demonstrate steering for
\begin{equation}
  \zeta\geq(d-1)^2\left[1-\left(1-\frac1d\right)^{1/(d-1)}\right].
  \label{eq:dicho-wer}
\end{equation}
This approximately evaluates to $0.7340$ for $d=3$ and $0.8230$ for $d=4$.
By direct comparison with values obtained in Table~\ref{tab:marvel} in the main text, the three indicated assemblages can easily be identified.
Two comments regarding the table are in order.
Firstly, notice that all of the three identified cases are concerning quantum steering of the Werner states.
While it was also proven~\cite{Nguyen2019b} that there are also finite measurement assemblages that can reveal quantum steering of the isotropic states which are unsteerable for all dichotomic measurements, such assemblages are not yet found with our construction.
Secondly, even for the Werner states, while the indicated assemblages perform better than all dichotomic measurements in demonstrate quantum steering, they are obviously strictly weaker than all projective measurements~\cite{Wiseman2007a,Nguyen2019b}; construction of better finite assemblages can therefore be expected in the future.

\subsection{The case of MUBs}

The fact that MUBs are uniform and rigidly symmetric is established later in Sec.~\ref{app:mub}.
With this property at hand we can use Eqs~\eqref{eq:tight-bound} and \eqref{eq:tight-bound-beta} to get $\alpha^\ast$ and $\beta^\ast$ (see Table~\ref{tab:clifford}).
While the value of $\alpha^\ast$ was already known~\cite{DSFB19}, the one of $\beta^\ast=1-\mu d/\abs{M}$ is new and has interesting consequences.
It was indeed shown in Ref.~\cite[Appendix E 3 d]{DFK19} that for odd prime power dimensions one has $\mu=0$ for MUBs.
With our results, and in particular Eq.~\eqref{eq:beta-proj}, this means that in these dimensions $\beta^\ast=1$, i.e., MUBs cannot be used to steer the Werner states.
A special case of this phenomenon in dimension $d=3$ had been pointed out in Ref.~\cite{Skrzypczyk2014a}.

\begin{table}[h]
  \centering
  \begin{tabular}{|c|c|c|c|c|c|c|}
    \hline
    $~d~$ & Isotropic state                                        & Werner state                                          \\ \hline
    2     & \multicolumn{2}{c|}{$\frac{1}{\sqrt{3}}\approx0.5774$}                                                         \\ \hline
    3     & $\quad\frac{1+3\sqrt{5}}{16}\approx0.4818\quad$        & 1                                                     \\ \hline
    4     & $\frac{3+2\sqrt{3}}{15}\approx0.4309$                  & $\frac{\sqrt{5}+\sqrt{10-2\sqrt{5}}}{5}\approx0.9174$ \\ \hline
    5     & $\approx0.3863$                                        & 1                                                     \\ \hline
    7     & $\approx0.3318$                                        & 1                                                     \\ \hline
    8     & $\frac{3+2\sqrt{3}}{21}\approx0.3078$                  & $\approx0.9981$                                       \\ \hline
    9     & $\approx0.2862$                                        & 1                                                     \\ \hline
    16    & $\gtrapprox0.2165$                                     & $\gtrapprox0.9997$                                    \\ \hline
    32    & $\gtrapprox0.1328$                                     & $\gtrapprox0.999993$                                  \\ \hline
  \end{tabular}
  \caption{
    Quantum steering with MUBs in prime power dimension $d$.
    Importantly, in odd prime power dimensions, quantum steering of the Werner states can never be revealed by using MUBs while this seems to be only asymptotically the case for even prime power dimensions.
  }
  \label{tab:clifford}
\end{table}

\section{Symmetry of MUBs}
\label{app:mub}

We have sketched the idea of the proof of the uniform and rigid symmetry of MUBs for odd prime dimensions in the main text.
In this section, we are going to give the detailed proofs for all odd and even prime power dimensions.
We start with the description of the construction of MUBs via the finite field phase space.
To make it self-contained, we also include all necessary materials on finite fields and Wigner functions; readers who are familiar with these concepts can skip the corresponding sections.
We then analyse the symmetry of MUBs described by the (galoisian) Clifford groups and thereby establish the uniformity and rigidity of their symmetry.
As the structures of the (galoisian) Clifford groups are different in odd and even dimensions,  the proofs differ in these two cases.

\subsection{Finite fields}

One may be familiar with the fact that for $p$ being a prime number, the set of residue classes modulo $p$ forms a field with the natural addition and multiplication, denoted $\FF_p$.
As for $d$ being a prime power dimension, $d=p^n$ for some positive integer number $n$, one can construct a so-called field extension of degree $n$ over $\FF_p$~\cite{Durbin2009a}.
Formally, from an irreducible polynomial $q(x)$ of degree $n$ in the ring $\FF_p[x]$ of all polynomials with coefficients in $\FF_p$, one forms the residue class ring $\FF_p [x]/\mean{q(x)}$, where $\mean{q(x)}$ is the ideal in $\FF_p [x]$ generated by $q(x)$.
For $q(x)$ being an irreducible polynomial, the residue class ring $\FF_p [x]/\mean{q(x)}$ is in fact a field, which is denoted $\FF_d$~\cite{Durbin2009a}.

Admittedly, the above formal construction may appear too abstract at first sight.
Conveniently, in practice, one only needs to work with derived properties of the finite field $\FF_d$, which are described below.
For a more extensive introduction, readers can consult Ref.~\cite{Woo87,WF89,Gibbons2004a}.

The field $\FF_d$ contains the prime field $\FF_p$ as its smallest subfield.
The field theoretical trace maps an element $x$ of $\FF_d$ to an element of the prime subfield $\FF_p$, specifically,
\begin{equation}
  \tr (x) = x + x^p + x^{p^2} + \cdots + x^{p^{n-1}}.
\end{equation}
Note that we use $\tr$ to denote the field theoretical trace, to be distinguished with the matrix trace $\Tr$.

With $\omega= \me^{2\i\pi/p}$, one can show that
\begin{equation}
  \frac{1}{d} \sum_{y \in \FF_d} \omega^{\tr (x y)} = \delta_{x,0}.
\end{equation}
The last equality allows one to perform a Fourier transform over functions on $\FF_d$, which looks very much like the normal discrete Fourier transform.

It is also sometimes helpful to think of the field $\FF_d$ as an $n$-dimensional vector space over $\FF_p$ (with an extra multiplicative structure).
Indeed, one can specify a basis $\{e_r: r=1,2,\ldots,n\}$ for $\FF_d$ such that any element $x$ can be written as
\begin{equation}
  x = \sum_{r=1}^n x_r e_r,
\end{equation}
with $x_i \in \FF_p$.
There exists a unique dual basis $\{\bar{e}_s: s=1,2,\ldots,n\}$ of $\FF_d$ such that
\begin{equation}
  \tr (e_r \bar{e}_s) = \delta_{rs}.
\end{equation}
Then one has $x_r = \tr (\bar{e}_r x)$.

\subsection{Displacement operators}
\label{app:displacement}

The presentation we use in this section closely follows Ref.~\cite{App09}.
Consider the Hilbert space of dimension $d=p^n$.
We choose a basis of the Hilbert space and label its elements with the finite field $\FF_d$, namely, $\{\ket{x}:x \in \FF_d\}$.
For each element $u\in\FF_d$, one defines
\begin{align}
  X_u \ket{x} &= \ket{x+u}, \\
  Z_u \ket{x} &= \omega^{\tr (ux)} \ket{x},
\end{align}
where $\omega=\me^{2\i\pi/d}$.
Then for $\u=(u_1,u_2) \in \FF_d^2$, one defines the \emph{displacement operator} to be
\begin{equation}
  D_{\u} = \tau^{\tr{(u_1 u_2)}} X_{u_1} Z_{u_2},
\end{equation}
where $\tau=\omega^{(p+1)/2}$.
Note that the map $D: \FF_d^2 \to U(d)$ is a projective representation of the linear translation group (i.e., the additive group) $\FF_d^2$.
In fact,
\begin{equation}
  D_{\u} D_{\v} = \tau^{\dprod{\u}{\v}} D_{\u + \v},
\end{equation}
where $\dprod{\u}{\v}$ is the standard symplectic form,
\begin{equation}
  \dprod{\u}{\v}= \tr (u_2 v_1 - u_1 v_2).
  \label{eq:symplectic_form}
\end{equation}

\subsection{Standard construction of MUBs and its freedom}

In this section, we describe the construction of MUBs by means of the geometry of the finite phase space $\FF_d^2$.
The set $\FF_d^2$ has the natural structure of an affine plane, where each vector $\u \in \FF_d^2$ can be regarded as a \emph{point}.
\emph{Lines} of $\FF_d^2$ are sets of the form $l=\{\u \in \FF_d^2: a u_1 + b u_2 = c\} $ for some $a,b,c \in \FF_d$.
Using the field structure of $\FF_d$, one can verify that through two points there is exactly one line, and that two lines can be either parallel or meet exactly at one point.
Lines that go through the origin, $l=\{x \u: x \in \FF_d\}$ for some $\u  \in\FF_d^2$,  are also called \emph{rays}.
There are exactly $d+1$ such rays.
See Ref.~\cite{Woo87,WF89,Gibbons2004a} for more discussions.

Considering the ray $l=\{x \u: x \in \FF_d\}$, the subgroup of $d$ displacement operators $\{ D_{\v}: \v \in l \}$ are clearly commuting.
These operators define a basis for the Hilbert space, or in other words, a projective measurement.
The bases corresponding to the $d+1$ different rays form MUBs~\cite{Gibbons2004a}.
The fact that the overlaps between effects of different measurements satisfy the unbiasedness condition (i.e., equal $1/\sqrt{d}$) follows directly from Ref.~\cite{Bandyopadhyay2002a} as clearly discussed in Ref.~\cite{Gibbons2004a}.

So far, each basis is associated to a ray of the finite plane $\FF_d^2$.
For each ray, there are exactly $d-1$ lines that are parallel to it.
In total, those $d$ lines cover the whole plane $\FF_d^2$ and are thus called a \emph{striation}~\cite{Gibbons2004a}.
One can further associate each element of a basis to a line in the striation defined by the ray in a way that manifests the symmetry of the assemblage.

Let $\mathscr{L}(\FF^2_d)$ denote the set of all lines in $\FF_d^2$ and let $Q: \mathscr{L}(\FF^2_d) \to M_d^+(\CC)$ be an association of a line of $\FF_d^2$ to a projection onto a vector of one of the MUBs as constructed above, which is called a quantum net in Ref.~\cite{Gibbons2004a}.
We demand that $Q$ is covariant under the action of the translation group, namely,
\begin{equation}
  Q(\u + l) = D_{\u} Q (l) D_{\u}^{-1} \mbox{ for all $l \in \mathscr{L} (\FF_d^2)$.}
\end{equation}
Any line can be reached by translating a ray with an appropriate translation.
It is then clear that $Q$ is completely fixed by the value it takes on rays.
Requiring $Q$ to manifest the symmetry of the assemblage described by the translation operator is nonetheless not sufficient to fix $Q$~\cite{Gibbons2004a}.
However MUBs have a higher symmetry group, namely the Clifford group, which we will discuss below.
In odd prime power dimensions, requiring $Q$ to manifest the symmetry described by the Clifford group will fix the choice of $Q$ (up to a specification of the computational basis).
In even prime power dimensions, the situation is different and we follow a different route.

\subsection{The galoisian Clifford group(s)}

It is to be noticed that there are different (equivalent and inequivalent) definitions of the Clifford group(s) in the literature.
Here we use the definition in the style of Ref.~\cite{App07,App09}.

Consider the group generated by all displacement operators with an arbitrary phase allowed, which is known as the (galoisian) Heisenberg--Weyl group $\mathrm{HW}(d)$,
\begin{equation}
  \mathrm{HW} (d) = \{\me^{\i\xi} D_{\u}:\u\in\FF^2_d,\xi\in\RR\}.
\end{equation}
Then one observes that the effects of MUBs are the projections onto the common eigenvectors of $d+1$ commutative subgroups of the Heisenberg--Weyl group~\cite{DEBZ10}.

Being interested in the symmetry of MUBs, we are looking for all unitary operators $U$ that map $\mathrm{HW}(d)$ to itself under conjugation.
This is known as its normaliser in the unitary group $\mathrm{U}(d)$, which is technically defined as the (galoisian) Clifford group~\cite{App09},
\begin{equation}
  C(d) = \{U \in \U(d): U \mathrm{HW}(d) U^{-1}= \mathrm{HW}(d) \}.
\end{equation}
Note that with this definition, the Clifford group $C(d)$ has infinite order.
However by quotienting out by the centre, which is the multiplication of the identity by an arbitrary phase, one obtains a finite group.
For our purposes, this quotienting is unnecessary, as the phase is automatically absorbed upon acting with conjugation.
In fact, we often only work with certain projective representation of a subset of $C(d)$.

Note that conjugation preserves commutativity.
Thus the Clifford group $C(d)$ also transforms commutative subgroups of $\mathrm{HW}(d)$ into each other without breaking them.
Thus the common eigenbases of the abelian subgroups are transformed into each other, without forming a new one.
In other words, $C(d)$ preserves the bundle projection, or MUBs are symmetric under the conjugate action of $C(d)$.

To demonstrate the uniformity and rigidity of the symmetry of MUBs, we look into the structure of the Clifford group.
It happens that the structure description of the Clifford group differs in odd power prime dimensions and in even power prime dimensions.
We thus discuss these two cases separately.

\subsection{Symmetry of MUBs: odd prime power dimensions}

\subsubsection{Representation of phase space transformations}

By $\SL(2,\FF_d)$, we denote the group of $2 \times 2$ matrices with elements in $\FF_d$ and unit determinant.
This is also the linear group that preserves the symplectic form~\eqref{eq:symplectic_form}.
We consider the group of transformations of the affine plane (transforming lines to lines) given by $\SL (2, \FF_d) \ltimes \FF_d^2$.
Recall that the semidirect product, denoted by $\ltimes$, is defined by the composition rule
\begin{equation}
  (F_1, \u_1) \circ (F_2, \u_2) = (F_1 F_2, \u_1 + F_1 \u_2),
\end{equation}
for $(F_1,F_2) \in \SL(2, \FF_d)$ and $\u_1,\u_2 \in \FF_d^2$~\cite{Armstrong2010a}.

In the following, one constructs a (projective) representation of $\SL (2, \FF_d) \ltimes \FF_d^2$ in odd prime power dimensions.
This forms a subgroup of the Clifford group $C(d)$, which allows us to select a particular ordering of the projections in each measurement of the MUBs, ordering which can be associated to each line in a way that manifests the symmetry of MUBs.

Although a faithful representation of $\SL(2,\FF_d) \ltimes \FF_d^2$ can be found, the matrix elements are somewhat cumbersome~\cite{App09}.
For our purposes, we can restrict ourselves to a projective representation $U: \SL(2,\FF_d) \ltimes \FF_d^2 \to U(d)$.
For any $F= \bigl(\begin{smallmatrix} \alpha & \beta \\ \gamma & \delta \end{smallmatrix}\bigr)$ in $\SL(2,\FF_d)$, one defines $U(F,\pmb{0})$ to be
\begin{equation}
  \left\{
    \begin{array}{ll}
      \frac{1}{\sqrt{d}} \sum_{x,y \in \FF_d} \tau^{\tr[(\alpha y^2 - 2 x y + \delta x^2)/\beta]} \ketbra{x}{y} & \mbox{if $\beta \ne 0$}, \\
      \sum_{x\in \FF_d} \tau^{\tr (\alpha \gamma x^2)} \ketbra{\alpha x}{x} & \mbox{if $\beta=0$}.
    \end{array}
  \right.
\end{equation}
Then one defines
\begin{equation}
  U(F,\v) = U(F,\pmb{0}) D_{\v}.
\end{equation}

Most importantly, we are interested in its conjugate action on the displacement operators.
For $(F,\v) \in \SL (2, \FF_d) \ltimes \FF_d^2$, we have
\begin{equation}
  U(F,\v) D_{\u} U(F,\v)^{-1} = \omega^{\dprod{\u}{F \v}} D_{F \u}.
  \label{eq:sl_covariant}
\end{equation}
It is then clear that the image of $U$ forms a subset of the Clifford group $C(d)$.
In fact, in the literature, $\SL(2,\FF_d) \ltimes \FF_d^2$ (or the group generated by its image) is known as the restricted (galoisian) Clifford group~\cite{App09}.

Recall that each striation is associated with on of the mutually unbiased bases.
But as one tries to associate each line in a striation with a projection onto one of the vectors of the basis, there is an ambiguity in choosing a vector in the basis to associate to the ray (line that goes through the origin) of the striation.
If one demands that the association of the projections with the rays has to be covariant under the action of $\SL(2,\FF_d)$, this ambiguity is resolved (up to the specification of the computational basis).
More precisely, one starts with associating the vertical axis $l_0$ to $P_0 = \ketbra{0}{0}$, where $\ket{0}$ is the $0$th state of the computational basis.
For $F \in \SL(2,\FF_d)$, the line $F l_0$ is associated with $U(F,\pmb{0}) P_0 U^{-1}(F,\pmb{0})$.
The ambiguity in choosing the projection for all other rays is thus resolved.
More importantly, the symmetry of MUBs by the restricted Clifford group can be studied by investigating the action of the group $\SL (2, \FF_d) \ltimes \FF_d^2$ on the lines of $\FF_d^2$, making all further arguments rather straightforward.
This happy situation does not happen for even prime power dimensions and one has to rely on a different approach.

\subsubsection{Uniformity}

The uniformity of MUBs follows directly from the above construction (and the procedure to fix the order of effects).
Indeed, one can easily check that the group $\SL (2, \FF_d) \ltimes \FF_d^2$ for odd prime power dimensions acts transitively on the lines, mapping any effect of the MUBs into any other.

\subsubsection{Rigidity}

The fact that the action of $\SL (2, \FF_d) \ltimes \FF_d^2$ on the lines of $\FF_d^2$ faithfully represents the symmetry of MUBs allows us to prove the rigidity of MUBs by studying the affine plane $\FF_d^2$.
For this to be carried out easily, we are to map also general operators to functions over $\FF_d^2$.
This gives rise to the notion of Wigner function over the finite field phase space~\cite{Woo87,Gibbons2004a}.

It is straightforward to verify that $\{D_{\u}: \u \in \FF_d^2 \}$ forms an orthonormal basis for the operator space, since $\Tr(D_{\v}^\dagger D_{\u}) = d \delta_{\u,\v}$~\cite{App09} .
This allows one to expand any operator as
\begin{equation}
  X= \frac{1}{d} \sum_{\u \in \FF_d^2} C_X(\u) D_{\u},
\end{equation}
where $C_X(\u) = \Tr (D_{\u}^\dagger X)$.
The function $C_X(\u)$ is known as the characteristic function of $X$.

Note that we are eventually interested in the subspace of hermitian operators.
As $D_{\u}$ are generally not hermitian, even when $X$ is hermitian, the characteristic function $C_X(\u)$ can take complex values in general.
To avoid this complex representation, one makes a Fourier transform to obtain the Wigner function,
\begin{equation}
  W_X(\u) = \sum_{\v \in \FF^2_d} \omega^{\dprod{\u}{\v}} C_X(\v).
\end{equation}
It is straightforward to show that as $X$ is hermitian, $W_X$ is real and as $X$ has unit trace, $\sum_{\v \in \FF_d^2} W_X(\u)=1$.
The properties of $W_X$ are in fact a lot like the Wigner function as defined for continuous variable as remarked in Refs~\cite{Woo87,Gibbons2004a}.
It is also straightforward to verify that for $(F,\v) \in \SL(2,\FF_d) \ltimes \FF_d^2$, one finds
\begin{equation}
  W_{U(F,\v) X U{(F,\v)}^{-1}} (\u) = W_X(\v + F\u).
\end{equation}

Let us come back to the rigidity of MUBs.
For the sake of specificity, we consider the vertical line through the origin $l_0$, which corresponds to $P_0$ (any line would work because the assemblage is uniform); see again Fig.~\ref{sfig:2c} in the main text.
Let us consider its stabiliser group.
It is easy to see that the stabiliser group contains all translations parallel to $l_0$.
Thus all points on vertical lines are in the same orbit.
Moreover, the stabiliser group must also contain the linear transformations of the form $\bigl(\begin{smallmatrix} \theta & 0 \\ 0 & \theta^{-1} \end{smallmatrix}\bigr)$, where $\theta$ is a primitive element of $\FF_d$ (that is, an element that generates the multiplicative group of $\FF_d$).
This shows that all points of the horizontal line going through the origin, except for the origin itself, are in the same orbit.
It is then clear that the stabiliser group acting on the phase space generates exactly two orbits: the line itself and its complement (see also the argument and Fig.~\ref{sfig:2c} in the main text).

Now, suppose that $X$ is invariant under the action of this subgroup of the Clifford group in the operator space, then its Wigner function $W_X$ is invariant under the action of $\SL(2,\FF_d) \ltimes \FF_d^2$ on the phase space $\FF_d^2$.
As a result, the Wigner function $W_X$ can only accept constant values on the orbits, which implies that it is the convex combination of the indicator function of the line and the constant function (everywhere).
Translated back to the operator space, this implies that the only proper projections that commute with the stabiliser group are $P_0$ or its complement.
This demonstrates that MUBs are rigidly symmetric in odd prime power dimensions.

\subsection{Symmetry of MUBs: even prime power dimensions}

\subsubsection{Multi-qubit representation and the generators of the Clifford group}

In this case, we consider the explicit realisation of the Hilbert space of the system as the tensor product of $n$ qubits.
To this end, we choose a basis $\{e_r: r=1,2,\ldots,n\}$ for the finite field $\FF_d$ ($d=2^n$).
The basis allows one to identify $x \in \FF_d$ with a string of binary letters, $(x_1,x_2,\ldots,x_n)$, $x_r \in \ZZ_2$, via the expansion
\begin{equation}
  x = \sum_{r=1}^n x_r e_r.
\end{equation}
Then the map
\begin{equation}
  S \ket{x} = \ket{x_1} \otimes \cdots \ket{x_n}
\end{equation}
establishes an isomorphism between the Hilbert space of dimension $d=2^n$ under consideration and the tensor product space of $n$ qubits~\cite{App09}.

Let $\u$ be a vector of $F_d^2$.
Then by expanding $u_1= \sum_{r=1}^{n} q_r e_r $ and $u_2= \sum_{r=1}^{n} p_r e_r$, we define $n$ vectors of $Z_2^2$, $\v_r=(q_r,p_r)$.
For each vector $\v_r$ of $Z_2^2$, let $D^{(2)}_{\v_r}$ denote the displacement operator acting on the qubit space $\CC^2$ as defined in Section~\ref{app:displacement}.
It is straightforward to show that~\cite{App09}
\begin{equation}
  S D_{\u} S^{-1} = D_{\v_1}^{(2)} \otimes D_{\v_2}^{(2)} \otimes \cdots \otimes D_{\v_n}^{(2)}.
\end{equation}
This identifies the (galoisian) Heisenberg--Weyl group $\mathrm{HW} (d)$ and the tensor product of the Heisenberg--Weyl groups defined on each qubit~\cite{App09}.
When distinguishing will be necessary for the sake of clarity, the latter will be called the multi-qubit Heisenberg--Weyl group.
Likewise, the (galoisian) Clifford group is identified with the multi-qubit Clifford group as the normaliser of the multi-qubit Heisenberg--Weyl group in the (global) unitary group.
It is well-known that the (multi-qubit) Clifford group is generated by the single-qubit Hadamard gates $H_j = \bigl( \begin{smallmatrix}1 & 1 \\ 1 & -1 \end{smallmatrix}\bigr)$, the single-qubit phase gates $P_j = \bigl(\begin{smallmatrix} 1 & 0 \\ 0 & i \end{smallmatrix}\bigr)$ (acting on qubit $j$) and the two-qubit CNOT gates $\mathrm{CNOT}_{jk} = \ketbra{0}{0} \otimes \I + \ketbra{1}{1} \otimes X$ (acting on the pair of qubits $(j,k)$), together with an irrelevant arbitrary phase~\cite{Gottesman1998a}, that is
\begin{equation}\nonumber
  C(2^n)=\left\langle \me^{\i\xi }, H_j, P_j, \mathrm{CNOT}_{jk}: \xi \in \RR, 1 \le j < k \le n \right\rangle.
\end{equation}

\subsubsection{Uniformity}

The uniformity of MUBs follows directly from results of Ref.~\cite{Wootters2007a}.
Indeed, the authors of Ref.~\cite{Wootters2007a} have proved a stronger property: there is a single element of the Clifford group that cycles over MUBs; see also~\cite{App09}.

\subsubsection{Rigidity}
To see the rigidity of MUBs, we pick up a projection and identify its stabiliser group.
By uniformity we can simply consider the projection onto the basic computational state, $P_0 = \ketbra{0,0,\ldots,0}{0,0,\ldots,0}$.
We see that the stabiliser contains at least all the phase gates $P_j$ and all the $\mathrm{CNOT}_{jk}$ gates,
\begin{equation}
  G_0= \left\langle  P_j, \mathrm{CNOT}_{jk}: \xi \in \RR, 1 \le j < k \le n \right\rangle.
\end{equation}
Suppose that a projection $\Pi$ commutes with $G_0$.
Note that the set of all phase gates are simultaneously diagonal in the computational basis $\{\ket{s}: s \in \{0,1\}^n \}$ with distinct sets of eigenvalues.
Thus if $\Pi$ commutes with all phase gates, it can only be of the form
\begin{equation}
  \Pi = \sum_{s \in S} \ketbra{s}{s},
\end{equation}
for some subset $S$ of binary strings of length $n$, i.e., some subset of $\{0,1\}^n$.
It is then easy to see that whenever $S$ contains a string differing from $(0,0,\ldots,0)$, by conjugating with an appropriate combination of $\mathrm{CNOT}_{ij}$, one can show that $\Pi$ contains the string $(1,1,\ldots,1)$.
Reversely, whenever $S$ contains the string $(1,1,\ldots,1)$, by conjugating with an appropriate combination of $\mathrm{CNOT}_{ij}$, one can show that $\Pi$ contains all other strings which differ from $(0,0,\ldots,0)$.
In combination, we see that once a string differing from $(0,0,\ldots,0)$ is contained in $S$, all other strings differing from $(0,0,\ldots,0)$ are also contained in $S$.
If $S$ also contains $(0,0,\ldots,0)$, $\Pi$ is trivially the identity operator, else it is precisely $\Pi=\I - P_0$.
This therefore establishes the rigidity of MUBs in even prime power dimensions.

\section{The \texttt{SQMA} package}
\label{app:sqma}

The package \texttt{SQMA} (Symmetry of Quantum Measurement Assemblage) under construction contains the code and the data for the constructed measurement assemblages in Section~\ref{app:construction}.
It will also include  implementation of the simplification of the SDP as discussed in~\ref{app:simplification}.
Commands to construct and work with MUBs and Clifford group(s) will also be available.
To exploit computational implementations with groups, \texttt{SQMA} is mainly written in \texttt{GAP}~\cite{GAP4}, and also makes use of \texttt{CAP}~\cite{CAP}, a package that implements computational category for \texttt{GAP}.
The complex reflection groups are imported from the package \texttt{CHEVIE}~\cite{GH96} in \texttt{GAP3}~\cite{GAP3}.
Interfaces with \texttt{Mathematica} and \texttt{Matlab} will also be provided.
We refer to the future github repository:
\begin{center}
  \texttt{https://gitlab.com/cn611340/sqma/}
\end{center}
for more detailed instructions.

\end{document}